\documentclass[12pt]{iopart}
\usepackage{iopams}

\expandafter\let\csname equation*\endcsname\relax

\expandafter\let\csname endequation*\endcsname\relax

\usepackage{amsmath,bm}
\usepackage{amsthm}
\usepackage{amssymb}
\usepackage{graphicx}

\usepackage{hhline}
\usepackage{amssymb}
\usepackage{amsthm}
\usepackage{mathrsfs}
\usepackage{braket}
\usepackage{caption}
\usepackage[colorlinks,citecolor=blue,urlcolor=blue,linkcolor=blue]{hyperref}
\usepackage{xcolor}
\usepackage{soul}
\usepackage{caption}
\usepackage{subcaption}

\newcommand{\lie}{\mathcal{L}}

\begin{document}

\title{Numerical solutions for the $f(R)$-Klein-Gordon system}

\author{Ulrich K. Beckering Vinckers$^{1,2,*}$, \'Alvaro de la Cruz-Dombriz$^{1,3}$ and Denis Pollney$^4$}
\address{$^1$ Cosmology and Gravity Group, Department of Mathematics and Applied Mathematics,
University of Cape Town, Rondebosch 7701, Cape Town, South Africa}
\address{$^2$ Van Swinderen Institute, University of Groningen, 9747 AG Groningen, The Netherlands}
\address{$^3$ Departamento de F\'isica Fundamental, Universidad de Salamanca,
    P. de la Merced, 37008 Salamanca, Spain}
\address{$^4$ Department of Mathematics, Rhodes University, Makhanda 6140, South Africa}

\eads{$^*$ bckulr002@myuct.ac.za}

\begin{abstract}
We construct a numerical relativity code based on the Baumgarte-Shapiro-Shibata-Nakamura (BSSN) formulation for the gravitational quadratic $f(R)$ Starobinsky model. By removing the assumption that the determinant of the conformal 3-metric is unity, we first generalize the BSSN formulation for general $f(R)$ gravity theories in the metric formalism to accommodate arbitrary coordinates for the first time. We then describe the implementation of this formalism to the paradigmatic Starobinsky model. We apply the implementation to three scenarios: the Schwarzschild black hole solution, flat space with non-trivial gauge dynamics, and a massless Klein-Gordon scalar field. In each case, long-term stability and second-order convergence is demonstrated. The case of the massless Klein-Gordon scalar field is used to exercise the additional terms and variables resulting from the $f(R)$ contributions. For this model, we show for the first time that additional damped oscillations arise in the subcritical regime as the system approaches a stable configuration. 
\end{abstract}

\section{Introduction}

One of the most widely studied modifications of classical General Relativity (GR) are the scalar tensor $f(R)$ theories, in which the Einstein-Hilbert gravitational action is generalized to be a nonlinear function of the Ricci scalar. While considerable progress has been made in understanding the implications of such theories~\cite{Sotiriou:2008rp,DeFelice:2010aj,Faraoni:2010pgm,Nojiri:2010wj,Nojiri:2017ncd}, the technical nature of these theories limits analytical understanding to systems with high-symmetry or perturbations of such systems. In recent years, numerical relativity has contributed greatly to our knowledge of the dynamics of space-time in classical GR, and provides some hope of having similar applicability to questions in $f(R)$ gravity.

The Baumgarte-Shapiro-Shibata-Nakamura (BSSN) formalism \cite{shibata,Baumgarte:1998te} is a modification of the Arnowitt-Deser-Misner (ADM) formulation \cite{Arnowitt:1959ah} which has become one of the most widely used formulations in the development of numerical relativity codes due to its high level of empirical stability and robustness \cite{alcubierre}. A corresponding BSSN formulation for $f(R)$ gravity has also been shown to be strongly hyperbolic in certain regimes and may prove useful as a basis for numerical implementation~\cite{Mongwane:2016qtz, Cao:2021mvy}.

The formulation presented in~\cite{Mongwane:2016qtz} is written with Cartesian coordinates in mind, which can lead to complications when the equations are applied, for instance, to problems with a radial coordinate~\cite{alcubierre,brown}. A main source of difficulty is the assumption that the conformal 3-metric has a unit determinant in Cartesian coordinates, and thus incompatible with the natural 3-metric developed using spherical polar coordinates. Another difficulty that arises is the singular nature of the initial data for the conformal connection function in spherical polar coordinates. Nevertheless, it has been shown in the context of GR that this difficulty may be treated by introducing a redefined field~\cite{alcubierre,brown,brown2,Montero:2012yr,Baumgarte:2012xy,Akbarian:2015oaa} whose initial data is zero.

Here, we present a modification of the BSSN formulation for $f(R)$ gravity developed in~\cite{Mongwane:2016qtz}, which allows for the determinant of the conformal 3-metric to be a general function rather than unity. Such a modification of the BSSN formulation is referred to as the ``generalized-BSSN'' (GBSSN) formulation and was constructed for the case of GR in \cite{brown,brown2} and examined further in \cite{alcubierre}. We have implemented the formulation in a numerical code, to demonstrate its potential for the study of scalar tensor $f(R)$ theories in the strong-field regime. In particular, we focus our attention on the quadratic Starobinsky model~\cite{Starobinsky:1979ty,Starobinsky:1980te,Vilenkin:1985md} of $f(R)$ gravity. This model has attracted increasing attention in recent years in the context of astrophysical applications, as well as having been widely used to provide a geometric origin to dark matter abundances \cite{Cembranos:2008gj}. It has also been proposed to explain cosmic inflation \cite{Starobinsky:1980te,Planck:2013jfk}. The modification to standard GR should be prominent in strong-gravity scenarios such as those studied in the present work.

For a specific numerical implementation, we study the Starobinsky-Klein-Gordon (SKG) system, i.e., the Starobinsky $f(R)$ gravity together with a matter action describing a massless Klein-Gordon (KG) scalar field. As tests of the robustness and accuracy of the code, we first apply it to two scenarios which are well-studied in the context of standard GR: $(i)$ the evolution of the Schwarzschild black hole solution (as has previously been done in~\cite{alcubierre,brown,Montero:2012yr,Baumgarte:2012xy}) using a fourth-order Runge-Kutta (RK4) scheme, and $(ii)$ a dynamical gauge pulse in a Minkowski space-time, as studied in \cite{alcubierre,Montero:2012yr}. While the standard set of GBSSN equations for GR together with the RK4 scheme do not allow for the stable evolution of a gauge pulse in flat space using spherical polar coordinates, it has been shown in \cite{alcubierre} that by regularizing certain evolution variables the system can be evolved stably. Alternatively, it was also shown in~\cite{Montero:2012yr} that stable evolutions are possible using a partially implicit Runge-Kutta (PIRK) scheme \cite{Cordero-Carrion:2012qac} without special regularizations. In the present work, we implement the latter and demonstrate the consistency and performance of our numerical approach through these test cases.

Finally, we apply the GBSSN formulation of $f(R)$ gravity to study the evolution of dynamical models in SKG gravity. We develop techniques for determining constraint-satisfying initial data for the geometry given a KG scalar field configuration, and demonstrate evolutions in these theories using the PIRK scheme. In particular, we note that the techniques used in \cite{alcubierre} to obtain initial data satisfying the Hamiltonian constraint are not suitable for the general $f(R)$ case. This is because the Hamiltonian constraint in the quadratic $f(R)$ case does not reduce to a linear differential equation and also contains higher-order spatial derivatives compared to the case of GR. We therefore require a different numerical approach to find suitable initial data. In addition, since the $f(R)$ GBSSN equations include additional evolution variables compared to GR, the numerical implementations used in \cite{alcubierre,Montero:2012yr,Akbarian:2015oaa} are not immediately applicable since we also need to perform the time integration of the additional evolution equations. Here, we develop for the first time a numerical implementation which accomodates these additional $f(R)$ GBSSN evolution variables. The results obtained when performing the evolution of the SKG system are then compared with those of the GR case, i.e., the Einstein-Klein-Gordon (EKG) system, which has been studied before using the GBSSN equations in \cite{alcubierre,Akbarian:2015oaa}.

This communication is organized as follows. Section~\ref{sec:f_R_review} reviews the field equations for a general $f(R)$ gravity theory within the metric formalism, i.e., when the metric tensor components are the only independent fields and the connection is the Levi-Civita one. In Section \ref{sec:foliations_of_space_time}, we review the $3+1$ decomposition of Riemannian space-time following the approach of~\cite{baumgarte_book,Baumgarte,Gourgoulhon}. In Sections~\ref{sec:ADM_f_R} and~\ref{sec:GBSSN_f_R} we modify the BSSN formulation of the $f(R)$ gravity given in~\cite{Mongwane:2016qtz} to accommodate curvilinear coordinates following \cite{alcubierre,brown,brown2} where this was done for GR. In Section~\ref{sec:reduction_to_spherical_symmetry} we consider the obtained system of evolution and constraint equations for the case of spherical symmetry. We emphasize that, up until this point, the discussion applies to general $f(R)$ theories. When subsequently performing numerical simulations in Section~\ref{sec:numerical_simulations} using these $f(R)$ GBSSN equations, we limit our consideration to the quadratic Starobinsky model. We consider a number of test cases, including a Schwarzschild black hole, a dynamical gauge pulse in flat space, and an SKG model whose results are compared with those of the EKG system. Some technical details regarding the PIRK scheme that we have implemented are presented in the Appendices for the interested reader. To this end, we discuss explicitly and for the first time how we have treated the additional GBSSN evolution variables that are present in the quadratic $f(R)$ case within the PIRK scheme in order to have stable and convergent numerical simulations.

\section{The $f(R)$ field equations}\label{sec:f_R_review}
We let $\mathcal{U}_4$ denote four-dimensional Riemannian space-time which, following \cite{hehl76,tseng}, is defined as the tuple $\left(\mathcal{M},\bm g,\nabla\right)$ where $\mathcal{M}$ is the space-time manifold, $\bm g$ is the metric tensor and $\nabla$ is the Levi-Civita connection. We consider a gravitational theory derived from $\mathcal{U}_4$ together with the total action
\begin{align}\label{eq:total_action}
S_{\text{total}}:=\frac{1}{16\pi}S_{\text G}+S_{\text m}\ ,
\end{align}
where\footnote{We note that the gravitational action \eqref{eq:gravitational_action} may be written as \cite{OHanlon:1972sdp,Wands:1993uu}
\begin{align}\label{eq:f_R_auxiliary_formulation}
S_{\rm G}\,=\,\int {\rm d}^4x\sqrt{-g}\left(\varphi R - V(\varphi)\right)\,,
\end{align}
where one defines $V(\varphi):=\xi(\varphi)\varphi-f(\xi(\varphi))$ and $\varphi:=f_R(\xi)$ where $\xi$ is an auxiliary field. In such a formulation, $\varphi$ describes the additional scalar degree of freedom that is present in general $f(R)$ gravity theories. In its form above, the  gravitational action should not be mistaken with the GR one, for which $\varphi=1$ and $V(\varphi)=0$. It is 
precisely due to the existence of $\varphi\neq1$, i.e., $f_R\neq1$, that the dynamics of the subsequent equations of motion differs from its GR counterparts. In particular, when working in the so-called metric formalism such equations become of higher (greater than two) order although the undesired Ostrogradski instability is avoided.
In the end, whether we consider the formulation \eqref{eq:f_R_auxiliary_formulation} or \eqref{eq:gravitational_action}, there will be one additional degree of freedom which needs to be accounted for in the subsequent $3+1$ decomposition; resulting in additional evolution equations compared to the case of GR.
}
\begin{align}\label{eq:gravitational_action}
S_{\text G}:=\int\text{d}^4x\sqrt{-g}\ f(R)\ ,
\end{align}
is the gravitational action and $S_{\text m}$ is some matter action \cite{Sotiriou:2008rp}. We note that here we use geometrized units with $G=c=1$ as well as the signature $\left(-,+,+,+\right)$. The Lagrangian density in the gravitational action \eqref{eq:gravitational_action} consists of a general function $f(R)$ of the Ricci scalar $R$. Varying the total action with respect to the metric tensor leads to the well-known equations of motion \cite{Sotiriou:2008rp}
\begin{align}\label{eq:f_R_field_equations}
f_RR_{\mu\nu}-\frac12fg_{\mu\nu}-\nabla_\mu\nabla_\nu f_R+g_{\mu\nu}\Box f_R=8\pi T_{\mu\nu}\ .
\end{align}
In the last expression, $\Box:=g^{\mu\nu}\nabla_\mu\nabla_\nu$ is the space-time covariant d'Alembertian operator and
\begin{align}\label{eq:energy_momentum_def_matter_action}
T_{\mu\nu}:=-\frac{2}{\sqrt{-g}}\frac{\delta S_{\text m}}{\delta g^{\mu\nu}}\ ,
\end{align}
is the energy-momentum tensor. In addition, we use $f_R:=\text df/\text dR$ and $f_{RR}:=\text d^2f/\text dR^2$\footnote{While it is common to denote derivatives of the $f(R)$ function with respect to the Ricci scalar using a prime, we will use a prime later on in Section \ref{sec:reduction_to_spherical_symmetry} to denote derivatives with respect to the radial coordinate.}. Taking the trace of the $f(R)$ field equations \eqref{eq:f_R_field_equations} yields
\begin{align}\label{eq:f_R_trace}
f_RR-2f+3\Box f_R=8\pi T\ ,
\end{align}
where $T:=T^\mu\phantom{}_\mu$ is the trace of the energy-momentum tensor.

\section{The 3+1 decomposition}\label{sec:foliations_of_space_time}

In order to re-cast the field equations as a system of evolution equations, we carry out a standard 3+1 decomposition of the four-dimensional space-time manifold $\mathcal{M}$ with metric $\bm g$ (see, for instance, \cite{Gourgoulhon,alcubierre2008introduction, baumgarte2010numerical}). Let $t$ be a non-constant scalar field on $\mathcal{M}$, and define the one-form $\text{\bf d}t$ normal to three-dimensional level surfaces of $t$, which we denote by $\Sigma_t$. We would like $t$ to serve as our time coordinate, and thus restrict our consideration to functions for which the dual of $\text{\bf d}t$ is timelike, so that $\Sigma_t$ is spacelike. We define
\begin{equation}
    \bm \Omega := \text{\bf d}t\,,
\end{equation}
and also define the normal vector field, $\bm n$, to the hypersurface $\Sigma_t$ through
\begin{align}
n^\mu=-\alpha g^{\mu\nu}\Omega_{\nu}\,.
\end{align}
The scalar $\alpha$ defined as
\begin{align}
\alpha^{-1}:=\sqrt{-\Omega^\mu\Omega_\mu}\,,
\end{align}
is called the \emph{lapse function} and gives the ratio between the proper time and coordinate time along integral curves of $t$. 

Given $\bm g$ on $\mathcal{M}$ we can induce a 3-metric, dubbed $\bm\gamma$, on each $\Sigma_t$ via
\begin{equation}
\gamma_{\mu\nu} = g_{\mu\nu} + n_\mu n_\nu\,.
\end{equation}
We use $D$ to denote the unique symmetric affine connection that is compatible with $\bm\gamma$, i.e., it satisfies
\begin{equation}
    D\bm \gamma = 0\,,
\end{equation}
and refer to this as the \emph{three-dimensional Levi-Civita connection}. For any tensor field $\bm B$ of type $\left(k,m\right)$, the three-dimensional covariant derivative $D\bm B$ is related to the space-time covariant derivative $\nabla \bm B$ through the projection into the hypersurface $\Sigma_t$,
\begin{equation}
    D_\alpha B^{\mu_1\dots \mu_k}{}_{\nu_1\dots\nu_m} =\gamma_\alpha{}^\beta\gamma_{\sigma_1}{}^{\mu_1}\dots \gamma_{\sigma_k}{}^{\mu_k}\gamma_{\nu_1}{}^{\rho_1}\dots \gamma_{\nu_m}{}^{\rho_m}\nabla_\beta{B}^{\sigma_1\dots \sigma_k}{}_{\rho_1\dots \rho_m}\,.
\end{equation}

The \emph{extrinsic curvature} of a slice $\Sigma_t$ is the symmetric 2-tensor
\begin{equation}\label{eq:deF_K}
    \bm K = -\frac{1}{2}\mathcal{L}_{\bm n}\bm \gamma\,.
\end{equation}
It can be shown that this is a purely spatial tensor satisfying $n^\mu K_{\mu\nu} = n^\nu K_{\mu\nu} = 0$. The \emph{intrinsic curvature} of a slice $\Sigma_t$ is described by the three-dimensional Riemann tensor, ${}^3\bm R$, which is defined through the linear map
\begin{equation}
    2D_{[\mu}D_{\nu]}w_\sigma={\!^3R}_{\mu\nu\sigma}\phantom{}^\rho w_\rho\,,
\end{equation}
together with the condition
\begin{equation}
    {\!^3R}_{\mu\nu\sigma\rho}{n}^\rho=0\,,
\end{equation}
which ensures that $\phantom{}^3\bm R$ lies entirely within $\Sigma_t$.

The space-time Riemann tensor $\bm R$ can be related to ${}^3\bm R$ through projections onto the spacelike slices. Specifically, one can find that
\begin{align}                       
    \gamma_\alpha{}^{\delta}\,\gamma_\beta{}^{\mu}\,\gamma_\sigma{}^{\nu}\,\gamma_\rho{}^{\lambda}{R}_{\delta\mu\nu\lambda} &= {^3\!{R}}_{\alpha\beta\sigma\rho} - {K}_{\alpha\rho}{K}_{\beta\sigma}+{K}_{\beta\rho}{K}_{\alpha\sigma}\,, \label{eq:gauss_equation}\\
    \gamma_\alpha{}^{\mu}\,\gamma_\beta{}^{\nu}\,\gamma_\sigma{}^{\lambda}\,n^\epsilon R_{\mu\nu\lambda\epsilon} &= D_\beta{K}_{\alpha\sigma} - D_\alpha{K}_{\beta\sigma}\,, \label{eq:codazzi_equation}\\
    \gamma_\mu^{\sigma}\,\gamma_\nu^{\rho}\,n^\alpha \,n^\beta{R}_{\alpha\sigma\rho\beta} &= -\mathcal{L}_{\bm{n}}K_{\mu\nu} - \frac{1}{\alpha}D_\mu D_\nu\alpha-K_\mu{}^{\beta}K_{\nu\beta}\,, \label{eq:ricci_equation}
\end{align}
which are the \emph{Gauss}, \emph{Codazzi} and $\emph{Ricci}$ equations, respectively. In deriving the last of these, we make use of the fact that the components of the 4-acceleration satisfy
\begin{align}\label{eq:deF_acceleration_vector}
    a_\mu:=n^\nu \nabla_\nu n_\mu = D_\mu \ln\alpha\,.
\end{align}
Equations~\eqref{eq:gauss_equation}-\eqref{eq:ricci_equation} encompass all of the non-trivial contractions of $\bm R$ onto the spacelike slices $\Sigma_t$ and thus give a complete description of the four-dimensional Riemann tensor in terms of $\bm n$ and quantities that can be calculated within a slice $\Sigma_t$ using $\bm\gamma$.

The Gauss, Codazzi and Ricci equations are sufficient to re-cast the four-dimensional Einstein field equations in terms of spatial quantities together with their derivatives along $\bm n$, as originally carried out in the ADM formulation of~\cite{Arnowitt:1959ah}. Similarly, they can be used to derive a 3+1 decomposition of the $f(R)$ field equations \eqref{eq:f_R_field_equations} which was carried out in \cite{Tsokaros:2013fma} and examined further in \cite{Mongwane:2016qtz}.

For the matter contribution, we follow the usual decomposition of the energy-momentum tensor as carried out in, for instance,~\cite{Gourgoulhon}. Given the definition of the energy-momentum tensor \eqref{eq:energy_momentum_def_matter_action}, the various projections onto the spacelike slices of the foliation determined by $\bm n$,
\begin{equation}
    \rho:=n^\mu n^\nu T_{\mu\nu}\,,\qquad 
    j^\mu:=-\gamma^{\mu\nu}n^\alpha T_{\nu\alpha}\,,\qquad
    S_{\mu\nu}:=\gamma_{\mu}^{\ \alpha}\gamma_{\nu}^{\ \beta}T_{\alpha\beta}\,,
\end{equation}
yield the \textit{energy density}, \textit{momentum density} and \textit{stress-energy tensor}, respectively.

\section{An ADM formulation for $f(R)$ theories}\label{sec:ADM_f_R}
In this section we briefly review an ADM-like formulation for $f(R)$ theories which was developed in~\cite{Mongwane:2016qtz,Tsokaros:2013fma}. In order to construct such a formulation, we make use of the $3+1$ decompositions of space-time and the energy-momentum tensor discussed in the previous section. A significant complication over standard GR results from the additional degrees of freedom embodied by the $f(R)\neq R$ part of the gravitational action \eqref{eq:gravitational_action}. The corresponding field equations~\eqref{eq:f_R_field_equations} involve second derivatives of $f$, i.e., fourth derivatives when expanded in terms of the metric. In fact, the field equations~\eqref{eq:f_R_field_equations} involve a d'Alembertian acting on $f_R$ which can be expanded as follows:
\begin{align}
    \Box f_R &= \gamma^{\mu\nu}\nabla_\mu\nabla_\nu f_R-n^\mu n^\nu\nabla_\mu\nabla_\nu f_R\label{eq:decompose_box_1}\\
    &= D^2f_R+K\lie_{\bm n}f_R-\lie^2_{\bm n}f_R+a^\nu D_\nu f_R\,.\label{eq:decompose_box}
\end{align}
As is done in~\cite{Mongwane:2016qtz,Tsokaros:2013fma}, we introduce an auxiliary variable based on the Lie derivative of the 4-Ricci scalar\footnote{We note that, while in \cite{Mongwane:2016qtz,Tsokaros:2013fma} the auxiliary field is defined as $\lie_{\bm n}R$, in the present work we have defined the auxiliary field \eqref{eq:W_definition} as $f_{RR}\lie_{\bm n}R$.}
\begin{equation}\label{eq:W_definition}
    W := \lie_{\bm n} f_R\,.
\end{equation}
Substituting this definition into the right-hand side of \eqref{eq:decompose_box} and substituting the trace \eqref{eq:f_R_trace} of the field equations into the left-hand side leads to the following result which can be interpreted as an evolution equation for $W$:
\begin{equation}\label{eq:evolution_W}
    \lie_{\bm n}W = \frac13\left(-8\pi T + f_RR-2f\right)+D^2 f_R + \gamma^{\mu\nu}D_\mu\ln\alpha D_\nu f_R+KW\,.
\end{equation}
Meanwhile, the definition~\eqref{eq:W_definition} itself can be re-cast as an evolution equation for the 4-Ricci scalar,
\begin{equation}\label{eq:evolution_Ricci}
\lie_{\bm n} R = \frac{W}{f_{RR}}\,,
\end{equation}
which is promoted to the status of an independent evolution variable provided $f_{RR}\neq0$.

\subsection{Evolution equations}
The geometrical variables associated with the 3+1 decomposition are the 3-metric, $\bm\gamma$, extrinsic curvature, $\bm K$, the 4-Ricci scalar, $R$, and the new variable $W$ defined in~\eqref{eq:W_definition}. Evolution equations for $W$ and $R$ can be derived from~\eqref{eq:evolution_W} and~\eqref{eq:evolution_Ricci}, respectively, while the evolution of the 3-metric is determined by the definition of the extrinsic curvature,~\eqref{eq:deF_K}, as usual. In order to derive an evolution equation for the extrinsic curvature, we first note that the field equations,~\eqref{eq:f_R_field_equations} and their trace~\eqref{eq:f_R_trace}, provide us with the following expression for the Ricci tensor
\begin{equation}\label{eq:f_R_in_terms_of_R_2}
R_{\mu\nu} = \frac{1}{f_R}\bigg[8\pi T_{\mu\nu}+\frac12fg_{\mu\nu}+\nabla_\mu \nabla_\nu f_R-\frac13g_{\mu\nu}\left(8\pi T+2f-f_RR\right)\bigg]\,.
\end{equation}
One can project this expression~\eqref{eq:f_R_in_terms_of_R_2} with reference to the 3+1 slicing to yield
\begin{align}
    n^\mu n^\nu R_{\mu\nu} &= \frac{1}{f_R}\left(8\pi\rho-\frac12f + D^2f_R\right) \label{eq:f_R_Ricci_n_n}\,,\\
    \gamma_\alpha{}^{\mu}n^\nu R_{\mu\nu} &= \frac{1}{f_R}\left(D_\alpha W+K_\alpha{}^{\nu}D_\nu f_R-8\pi j_\alpha\right) \label{eq:f_R_Ricci_n_g}\,,\\
    \gamma_\alpha{}^{\mu}\gamma_\beta{}^{\nu}R_{\mu\nu} &= \frac{1}{f_R}\left\{ 8\pi S_{\alpha\beta} + \frac12f\gamma_{\alpha\beta} + D_\alpha D_\beta f_R - \frac13\gamma_{\alpha\beta} \left[8\pi\left(S - \rho\right) + 2f-f_RR\right]\right\}\,.\label{eq:f_R_Ricci_g_g}
\end{align}
Substituting the last projected expression~\eqref{eq:f_R_Ricci_g_g} into the Ricci equation \eqref{eq:ricci_equation} yields an evolution equation for the extrinsic curvature tensor
\begin{align}\label{eq:adm_f_R_K_ij}
    \lie_{\bm n}K_{ij} &= \!^3R_{ij} - 2K_{il}K^l{}_j + KK_{ij}-\frac1{f_R} \bigg\{8\pi S_{ij} + \frac12 f\gamma_{ij} + D_iD_j f_R \nonumber \\
    & -\frac13\gamma_{ij}\left[8\pi\left(S -\rho\right) + 2f -f_RR\right]\bigg\} -\frac{1}{\alpha}D_iD_j\alpha\,.
\end{align}
In the last expression, we have used the convention of denoting spatial components using Latin indices. Finally, we transform the right-hand side of the evolution equations into derivatives along the natural time coordinate by defining the vector field
\begin{equation}
    \bm t = \alpha \bm n + \bm \beta\,,
\end{equation}
where $\bm\beta$ is spatial and is referred to as the \emph{shift vector}. In terms of the lapse, shift and 3-metric, the line-element associated with the full metric tensor is \cite{Baumgarte}
\begin{align}
\text{d}s^2=-\alpha^2\text{d}t^2+\gamma_{ij}\left(\text{d}x^i+\beta^i\text{d}t\right)\left(\text{d}x^j+\beta^j\text{d}t\right)\,.
\end{align}

In summary, the following set of equations provide a prescription for the time evolution of the geometrical variables:
\begin{align}
    \partial_t \gamma_{ij} &= -2\alpha K_{ij} + 2D_{(i}\beta_{j)}\,, \label{eq:g-evo}\\
    \partial_t K_{ij} &= \alpha\left(\!^3R_{ij} - 2K_{il}K^l{}_j + KK_{ij}\right) - \frac{\alpha}{f_R} \bigg[8\pi S_{ij} + \frac12 f\gamma_{ij} + D_iD_j f_R \nonumber \\
    & -\frac13\gamma_{ij}\left(8\pi\left(S -\rho\right) + 2f -f_RR\right)\bigg] - D_iD_j\alpha \nonumber\\
    & + 2K_{k(i}D_{j)}\beta^k + \beta^k D_k K_{ij}\,, \label{eq:K-evo}\\
    \partial_t W &= \frac13\left(-8\pi T + f_RR-2f\right)+D^2 f_R + \gamma^{ij}D_i\ln\alpha D_j f_R+KW + \beta^k D_k W\,,  \label{eq:evolve_W}\\
    \partial_t R &= \frac{\alpha W}{f_{RR}} + \beta^k D_k R\,. \label{eq:evolve_R}
\end{align}

\subsection{Constraints}
Having stated the evolution equations in the previous subsection, we now turn our attention to deriving the constraint equations. Contracting the Gauss equation~\eqref{eq:gauss_equation} twice leads to
\begin{equation}\label{eq:gauss_contracted}
    {}^3R+K^2-K_{ij}K^{ij} = R+2n^\mu n^\nu R_{\mu\nu}\,,
\end{equation}
into which we can substitute~\eqref{eq:f_R_Ricci_n_n} to give
\begin{equation}\label{eq:hamiltonian_constraint}
    \mathcal{H}:={}^3R+K^2 - K_{ij}K^{ij} - R - \frac{1}{f_R}\left(16\pi\rho - f + 2{D^2}f_R\right) = 0\,.
\end{equation}
This is the \emph{Hamiltonian constraint} for the $f(R)$ theory. Meanwhile, taking the trace of the Codazzi equation~\eqref{eq:codazzi_equation} and substituting in~\eqref{eq:f_R_Ricci_n_g} yields
\begin{equation}\label{eq:momentum_constraint}
    \mathcal{M}_{i}:=D_j K^j{}_i - D_i K - \frac{1}{f_R}\left(8\pi j_i - D_i W - K_i{}^j D_j f_R\right) = 0\,,
\end{equation}
dubbed as the corresponding \emph{momentum constraint}. As a consequence of the Bianchi identities, if these constraints \eqref{eq:hamiltonian_constraint} and \eqref{eq:momentum_constraint} are satisfied on an initial spacelike slice then they are preserved by the evolution equations.

\section{GBSSN evolution equations}\label{sec:GBSSN_f_R}

In the case of standard GR, a direct implementation of the evolution system contained in equations~\eqref{eq:g-evo} and \eqref{eq:K-evo} is known to lead to unstable numerical behaviour in $3+1$ simulations. A prescription which has been successful in curing this problem is to introduce well-chosen auxiliary dynamical variables. Such a prescription introduces additional degrees of freedom but has the effect of re-casting the partial differential equation system in a strongly hyperbolic form. In particular, the so-called BSSN system~\cite{shibata,Baumgarte:1998te} has proved to be a popular choice in 3+1 numerical relativity. In what follows, we briefly outline the BSSN formulation for $f(R)$ theories following \cite{Mongwane:2016qtz,Baumgarte} and extend this to accommodate arbitrary coordinates.

To begin, we introduced the \textit{conformal 3-metric}
\begin{align}\label{eq:conformal_spatial_metric_definition}
\bar{\bm\gamma}:= {\text e}^{4\phi}\bm{\gamma}\,,
\end{align}
where $\phi$ is the \emph{conformal factor} and is related to the determinants of the 3-metric and conformal 3-metric by
\begin{equation}\label{eq:phi_gam_bar_gam}
    \phi = -\frac{1}{12}\ln (\gamma/\bar{\gamma})\,.
\end{equation}
The connection coefficients of the corresponding conformal three-dimensional Levi-Civita connection $\bar D$, uniquely satisfying $\bar D\bar{\bm\gamma}=0$, are related to those of $D$ by
\begin{align}\label{eq:conformal_connection_coefficients}
  {^{3\,\,}\!\bar{\Gamma}}\phantom{}^k_{\ ij}={^{3\,\,}\!{\Gamma}}\phantom{}^k_{\ ij}+2\gamma^{kl}\left(\gamma_{lj}D_i\phi+\gamma_{li}D_j\phi-\gamma_{ij}D_l\phi\right)\,,
\end{align}
where $\gamma^{ij}$ denotes the inverse of the 3-metric tensor. 

The conformal 3-metric $\bar{\gamma}_{ij}$ and factor $\phi$ are the metric evolution variables of the new system. In standard BSSN, the additional degree of freedom resulting from the introduction of $\phi$ is fixed by placing a restriction on the determinant of the conformal metric, namely $\bar{\gamma}=1$. More generally~\cite{brown,brown2}, we can impose the evolution equation
\begin{equation}\label{eq:evolve_conformal_determinant}
    \left(\partial_t-\lie_{\bm\beta}\right)\ln\bar\gamma=-2vD_i\beta^i\,,
\end{equation}
where $v\in\{0,1\}$. Setting the parameter $v=0$ corresponds to the so-called \emph{Eulerian} condition while $v=1$ corresponds to the \emph{Lagrangian} description.

It is at this point that the formulation presented here differs from that in \cite{Mongwane:2016qtz}. More specifically, in \cite{Mongwane:2016qtz}, Cartesian coordinates are assumed and the determinant of the conformal 3-metric is set to unity. Here, we do not impose such a condition on $\bar\gamma$ and instead assign the evolution equation \eqref{eq:evolve_conformal_determinant}; considering arbitrary coordinates as a result. While this was done for the case of GR in \cite{brown,brown2}, and dubbed the GBSSN formulation, this has not be done before for a general $f(R)$ theory. In what follows we derive, for the first time, the GBSSN formulation for $f(R)$ gravity.

By making use of equations \eqref{eq:phi_gam_bar_gam} and \eqref{eq:evolve_conformal_determinant}, one finds the following evolution equation for the conformal factor \cite{brown}
\begin{align}\label{eq:evolution_equation_phi}
    \partial_t\phi=\frac{\alpha}6K-\frac{v}6D_k\beta^k+v\beta^k\partial_k\phi-\frac{v}6\beta^k\partial_k\ln\bar{\gamma}^{1/2}\ ,
\end{align}
where we have made use of the fact that
\begin{align}
{^{3}\!\phantom{}}\ \Gamma\phantom{}^k_{\ kj}={^{3}\!\phantom{}}\ {\bar{\Gamma}}\phantom{}^k_{\ kj}-6\ \partial_j\phi\,,
\end{align}
which follows from equation \eqref{eq:conformal_connection_coefficients}. An alternative to $\phi$ which we adopt here, and is similar to what is used in~\cite{brown,Campanelli:2005dd}, is to use the variable $\chi = e^{2\phi}$ which evolves according to
\begin{equation}\label{eq:evolution_equation_chi}
\partial_t\chi=\chi\left(\frac{\alpha}{3}K-\frac{v}3D_k\beta^k+\frac{v\beta^k\partial_k\chi}{\chi}-\frac{v}3\beta^k\partial_k\ln\bar\gamma^{1/2}\right)\,.
\end{equation}
Following the standard BSSN prescription, we decompose the extrinsic curvature tensor $\bm K$ into its trace, $K$, and trace-free part
\begin{align}\label{eq:decompose_extrinsic_curvature}
    \bm A:=\bm K-\frac13K\bm\gamma\,,
\end{align}
and carry out a conformal rescaling consistent with that of the 3-metric to define
\begin{align}\label{eq:rescaled_trace_free_definition}
\bar{\bm A}:={\text e}^{4\phi}\bm A\,.
\end{align}

An evolution equation for the conformal metric $\bar{\bm\gamma}$ follows by substituting its definition~\eqref{eq:conformal_spatial_metric_definition} into the evolution equation for the 3-metric~\eqref{eq:g-evo} and making note of~\eqref{eq:evolution_equation_phi}. After some algebra, we arrive at
\begin{align}\label{eq:evolution_conformal_spatial_metric}
\partial_t\bar{\gamma}_{ij}=-2\alpha\bar{A}_{ij}+2\bar{\gamma}_{k(i}\partial_{j)}\beta^k+\beta^k\partial_k\bar{\gamma}_{ij}+4\left(v-1\right)\bar{\gamma}_{ij}\beta^k\partial_k\phi\nonumber\\
-\frac23v\bar{\gamma}_{ij}\partial_k\beta^k-\frac23v\bar{\gamma}_{ij}\beta^k\partial_k\ln\bar{\gamma}^{1/2}\ .
\end{align}
Similarly, by substituting the definitions of the new variables into the evolution equation~\eqref{eq:K-evo} of the extrinsic curvature we find the following evolution equations for $K$ and the components of $\bar{\bm A}$:
\begin{align}\label{eq:evolution_trace_extrinsic}
    \partial_tK &= \alpha\left(\bar{A}_{ij}\bar{A}^{ij}+\frac{1}{3}K^2\right)+\beta^k\partial_kK-{D^2}\alpha+\frac{\alpha}{f_R}\left(8\pi\rho-\frac{f}{2}+{D^2}f_R\right)\,,\\
    \partial_t\bar{A}_{ij} &= -2\alpha\bar{A}_{ik}\bar{A}^k_{\ j}+\alpha K\bar{A}_{ij}-{\rm e}^{4\phi}\left(D_iD_j\alpha\right)^{\textup{TF}}+\alpha {\rm e}^{4\phi}\,{^{3\,}\!R}\phantom{}^{\textup{TF}}_{ij}+2\bar{A}_{k(i}\partial_{j)}\beta^k \nonumber\\
    & +\beta^k\partial_k\bar{A}_{ij}-\frac{2v}{3}\bar{A}_{ij}\partial_k\beta^k-\frac{2v}{3}\bar{A}_{ij}\beta^k\partial_k\ln\bar{\gamma}^{1/2}+4\left(v-1\right)\bar{A}_{ij}\beta^k\partial_k\phi \nonumber \\
    & -\frac{\alpha{\text e}^{4\phi}}{f_R}\left[8\pi S^{\text{TF}}_{ij}+\left(D_iD_j f_R\right)^{\text{TF}}\right]\,,\label{eq:evolution_equation_bar_A}
\end{align}
where the superscript $\text{TF}$ is used to indicate the trace-free part of the corresponding tensor field.

Finally, the GBSSN system introduces a further set of auxiliary variables given by components of the conformal Levi-Civita connection, namely
\begin{align}\label{eq:connection_function_definition}
    \phantom{}^3\bar\Gamma\phantom{}^i:=\bar\gamma\phantom{}^{jk}\phantom{}^3\bar\Gamma\phantom{}^i_{\ jk}
    =\partial_j\bar\gamma\phantom{}^{ij}-\bar\gamma\phantom{}^{ij}\partial_j\ln\bar\gamma^{1/2}\ .
\end{align}
Although they can be derived from the conformal 3-metric, these are treated as independent variables. An evolution equation follows from taking the time derivative of equation \eqref{eq:connection_function_definition} and substituting in equations \eqref{eq:momentum_constraint}, \eqref{eq:evolve_conformal_determinant} and \eqref{eq:evolution_conformal_spatial_metric}:
\begin{align}\label{eq:evolution_connection_function}
    \partial_t\ ^{3\,}\!{\bar{\Gamma}}\phantom{}^i& =2\alpha\ \!^3\phantom{}\bar\Gamma\phantom{}^i_{\ jk}\bar{A}^{jk}-\frac43\alpha\bar{\gamma}^{ik}\partial_kK-12\alpha\bar{A}^{ij}\partial_j\phi+4(1-v)\ \!^3{\bar{\Gamma}}\phantom{}^i\beta^k\partial_k\phi-\!^3{\bar{\Gamma}}\phantom{}^k\partial_k\beta^i \nonumber\\ 
    & -2\bar{A}^{ij}\partial_j\alpha+\beta^k\partial_k\!^3{\bar{\Gamma}}\phantom{}^i+\bar{\gamma}^{kj}\partial_k\partial_j\beta^i+2(1-v)\bar{\gamma}^{il}\partial_k\phi\partial_l\beta^k+2(1-v)\bar{\gamma}^{il}\beta^k\partial_l\partial_k\phi\nonumber\\
    & +\frac{v}{3}\bar{\gamma}^{ij}\left(\partial_j\partial_k\beta^k+\partial_l\ln\bar{\gamma}^{1/2}\partial_j\beta^l+\beta^k\partial_j\partial_k\ln\bar{\gamma}^{1/2}\right)+\frac{2v}{3}\ \!^3{\bar{\Gamma}}\phantom{}^i\beta^k\partial_k\ln\bar{\gamma}^{1/2}\nonumber\\
    &+\frac{2v}{3}\ \!^3{\bar{\Gamma}}\phantom{}^i\partial_k\beta^k-\frac{2\alpha\bar\gamma\phantom{}^{ik}}{f_R}\left(8\pi j_k-D_kW-\bar A\phantom{}_k^{\ j}D_j f_R-\frac13KD_k f_R\right)\ .
\end{align}
As noted in \cite{alcubierre}, when one imposes the Lagrangian condition in equation \eqref{eq:evolve_conformal_determinant}, i.e., setting $v=1$, the GBSSN formulation of GR reduces to the BSSN formulation when specifying Cartesian coordinates. Similarly, here we note that when using Cartesian coordinates, the GBSSN formulation presented above for $f(R)$ gravity reduces to the $f(R)$ BSSN formulation given in \cite{Mongwane:2016qtz} when we set $v=1$.
\section{Reduction to spherical symmetry}\label{sec:reduction_to_spherical_symmetry}
While the equations derived in the previous section apply to general geometries, it is also useful to consider the reduction to spherical symmetry both as a testing ground with fewer variables and simpler equations, as well as a platform for studying certain physical models such as irrotational collapse.
For this purpose, we re-write the GBSSN system in terms of standard spherical polar coordinates $(r,\theta,\psi)$. In spherical symmetry, the conformal 3-metric \eqref{eq:conformal_spatial_metric_definition} has only two independent components to be denoted as $a=a(t,r)$ and $b=b(t,r)$, and takes the diagonal form
\begin{align}\label{assumption1GBSSN}
    \bar{\gamma}_{ij} = \begin{pmatrix} 
        a & 0 & 0 \\
        0 & r^2b & 0 \\
        0 & 0 & r^2b\sin^2\theta \\
    \end{pmatrix} \,,
\end{align}
which was used in \cite{alcubierre}. The rescaled traceless extrinsic curvature $\bar{\bm A}$ has a single independent component, which we denote $A_a=A_a(t,r)$, and can be expressed as
\begin{align}\label{assumption2GBSSN}
    \bar{A}_i{}^j = A_a\begin{pmatrix} 
        1 & 0 & 0 \\
        0 & -1/2 & 0 \\
        0 & 0 & -1/2 
    \end{pmatrix} \ .
\end{align}
As suggested in \cite{brown}, this choice of variable ensures that $\bar{A}_{ij}$ is trace-free by construction. For the conformal connection function, a straightforward coordinatization derived from the metric is
\begin{align}\label{assumption3GBSSN}
    \!^3\phantom{}{\bar{\Gamma}}\phantom{}^i = \begin{pmatrix} 
    \!^3\phantom{}{\bar{\Gamma}}\phantom{}^r \\
    -\cos\theta/\left(r^2 b\sin\theta\right) \\
    0 \\
\end{pmatrix} \ .
\end{align}
However, this choice is awkward since its flat space initial data, ${}^3\bar{\Gamma}^r=-2/r$, is singular at the origin. Rather, following~\cite{alcubierre,brown2}, we introduce the alternative variable
\begin{equation}
    \bar{\Delta}^i := {}^3\bar{\Gamma}^i - \bar{\gamma}^{jk}\,{}^3\overset{\circ}{\Gamma}{}^i{}_{jk}\,,
\end{equation}
where ${}^3{\overset{\circ}\Gamma}{}^i{}_{jk}$ are the connection coefficients associated with the Minkowski metric. These have the advantages that they are components of a true vector and have initial data $\bar{\Delta}^r=0$ in flat space. In terms of the conformal 3-metric components $a$ and $b$, the regularized variable $\bar\Delta^r$ is \cite{alcubierre}
\begin{align}\label{eq:define_Delta_a_b}
    \bar\Delta^r={^{3\,}\!\phantom{}}{\bar\Gamma}\phantom{}^r+\frac{2}{rb} = \frac{a'}{2a^2} - \frac{b'}{ab} - \frac{2}{ra} + \frac{2}{rb}\,,
\end{align}
where the prime denotes differentiation with respect to the radial coordinate $r$.

\subsection{Evolution equations}
By imposing spherical symmetry and the Lagrangian condition ($v=1$), equations \eqref{eq:evolution_equation_chi} and \eqref{eq:evolution_conformal_spatial_metric} provide us with the following evolution equations for the metric variables:
\begin{align}
    \partial_t\chi &= 2\chi\left[\frac{\alpha K}{6}-\frac{\beta^r\phantom{}'}{6}+\frac{\beta^r\chi'}{2\chi}-\frac{\beta^r}{12}\left(\frac{a'}{a}+\frac4r+\frac{2b'}{b}\right)\right]\,,\label{eq:evolution_equation_chi_2}\\
    \partial_ta &= -2\alpha aA_a+\frac43a\beta^r\phantom{}'+\beta^ra'-\frac{1}3a\beta^r\left(\frac{a'}{a}+\frac4r+\frac{2b'}b\right)\,,\label{eq:evolution_a}\\
    \partial_tb &= \alpha bA_a-\frac{2}3b\beta^r\phantom{}'+\frac{2\beta^rb}r+\beta^rb'-\frac{1}3b\beta^r\left(\frac{a'}{a}+\frac4r+\frac{2b'}b\right)\,.\label{eq:evolution_b}
\end{align}
The two variables determining the extrinsic curvature, $K$ and $A_a$, evolve according to
\begin{align}\label{eq:evolution_trace_extrinsic_sph}
    \partial_tK &= -\frac{\chi^2}{a}\left[\alpha''-\alpha'\left(\frac{a'}{2a}+\frac{\chi'}{\chi}-\frac2r-\frac{b'}b\right)\right]+\alpha\left(\frac32A_a^2+\frac{K^2}3\right)+\beta^rK'\nonumber\\
    & +\frac{\alpha}{f_R}\left\{8\pi\rho-\frac{f}{2}+\frac{\chi^2}{a}\left[f_R''-f_R'\left(\frac{a'}{2a}+\frac{\chi'}{\chi}-\frac2r-\frac{b'}{b}\right)\right]\right\}\,,\\
    \partial_tA_a &= \alpha KA_a-\frac{\chi^2}{a}\left({D}_r{D}_r\alpha\right)^{\textup{TF}}+\frac{\alpha\chi^2}{a}\,{\!^3\phantom{}}R_{rr}^{\textup{TF}}+\beta^rA_a' \nonumber\\
    & -\frac{\alpha\chi^2}{af_R}\left[8\pi S^{\text{TF}}_{rr}+\left(D_rD_r f_R\right)^{\text{TF}}\right] \label{eq:evolution_A_a}\,,
\end{align}
where the $(r,r)$ component of the trace-free part of the Ricci tensor can be rewritten as
\begin{align}\label{eq:Ricci_TF}
    {\!^3\phantom{}}R_{rr}^{\textup{TF}} &= \frac23\bigg[-\frac{1}{r^2}+\frac{b''}{2b}+\frac{2b'}{rb}-\frac{3a'}{ar}-\frac{3b'a'}{2ab}-\frac{a''}{2a}+\frac{a'\phantom{}^2}{a^2}+a\left(\frac{1}{r^2b}+\frac{2b'}{rb^2}+\bar\Delta^r\phantom{}'\right)\nonumber\\
    & -\frac{\chi'}{\chi}\left(\frac2r+\frac{b'}{b}-\frac{a'}{a}\right)+\frac{\chi''}{\chi}-\frac{\chi'\phantom{}^2}{\chi^2}+2a'\left(\bar\Delta^r-\frac{a'}{2a^2}+\frac{b'}{ab}-\frac{2}{rb}+\frac{2}{ra}\right)\nonumber\\
    & -2\left(\frac{b'}{2b}-\frac{\chi'}{\chi}\right)^2+3\left(\frac{a'}{2a}-\frac{\chi'}{\chi}\right)\left(\frac1r+\frac{b'}{2b}-\frac{\chi'}{\chi}\right)-\frac4r\left(\frac{b'}{2b}-\frac{\chi'}{\chi}\right)\bigg]\,.
\end{align}
In computing the right-hand side of equation \eqref{eq:evolution_A_a}, it is useful to note that for any smooth function $E$, the following expansion holds
\begin{align}
    \left(\,{\!^3D}_r{\!^3D}_r E\right)^{\textup{TF}}=\frac23\left[E''-E'\left(\frac{a'}{2a}+\frac{b'}{2b}+\frac1r-\frac{2\chi'}{\chi}\right)\right]\ .
\end{align}
The use of the derivative of $\bar{\Delta}^r$ in the right-hand side of equation~\eqref{eq:Ricci_TF} has been shown to be important for the numerical stability of the evolution system~\cite{alcubierre,brown,Montero:2012yr,Akbarian:2015oaa}. Additionally,  as suggested in~\cite{alcubierre}, we also introduce the field $\bar\Delta^r$ itself in the second line of equation~\eqref{eq:Ricci_TF} in the form of an additional term which evaluates to zero in the case that the constraint~\eqref{eq:define_Delta_a_b} is satisfied exactly.

The evolution equation for $\bar{\Delta}^r$ follows from equations~\eqref{eq:evolution_connection_function},~\eqref{eq:define_Delta_a_b} and~\eqref{eq:evolution_b}:
\begin{align}\label{eq:evolution_Delta}
    \partial_t\bar{\Delta}^r &= \frac{2\alpha A_a}a\left(\frac{a'}{2a}+\frac{b'}{2b}+\frac1r\right)-\frac43\frac{\alpha K'}{a}-\frac{6\alpha A_a\chi'}{a\chi}-\left(\frac{a'}{2a^2}-\frac{b'}{ab}-\frac{2}{ar}\right)\beta^r\phantom{}'\nonumber\\
    & - \frac{2A_a\alpha'}a+\beta^r\left(\bar{\Delta}^r\phantom{}'+\frac{2}{r^2b}+\frac{2b'}{b^2r}\right)+\frac{\beta^r\phantom{}''}a+\frac{\beta^r}3\left(\frac{a'}{2a^2}-\frac{b'}{ab}-\frac{2}{ar}\right)\left(\frac{a'}a+\frac4r+\frac{2b'}b\right)\nonumber\\
    & + \frac{2\beta^r\phantom{}'}3\left(\frac{a'}{2a^2}-\frac{b'}{ab}-\frac{2}{ar}\right) +\frac{1}{3a}\Bigg[\beta^r\phantom{}''+\beta^r\phantom{}'\left(\frac{a'}{2a}+\frac2r+\frac{b'}b\right)\nonumber\\
    & + \beta^r\Bigg(\frac{a''}{2a}-\frac{a'^2}{2a^2}-\frac{2}{r^2}+\frac{b''}{b}-\frac{b'\phantom{}^2}{b^2}\Bigg)\Bigg]-\frac{2}{rb^2}\bigg[\alpha bA_a-\frac{2}3b\beta^r\phantom{}'+\frac{2\beta^rb}r+\beta^rb'\nonumber \\
    & - \frac{1}3b\beta^r\left(\frac{a'}{a}+\frac4r+\frac{2b'}b\right)\bigg]-\frac{2\alpha}{af_R}\left[8\pi j_r-W'-\left(A_a+\frac13K\right)f'_R\right]\,.
\end{align}
As noted above, in the case of $f(R)\neq R$ theories we employ the two additional variables $R$ and $W$. In spherical symmetry their evolution equations~\eqref{eq:evolve_W} and~\eqref{eq:evolve_R} reduce to
\begin{align}
    \partial_tW &= \frac{\alpha}{3}\left[Rf_R-2f-8\pi\left(S-\rho\right)\right]+\frac{\chi^2\alpha}{a}\left[f_R''-f_R'\left(\frac{a'}{2a}+\frac{\chi'}{\chi}-\frac{2}{r}-\frac{b'}{b}\right)\right]\nonumber\\
    &+ \frac{\chi^2\alpha'f_R'}{a}+\alpha KW+\beta^r\partial_rW\,,\label{eq:evolve_W_GBSSN}\\
    \partial_tR &= \frac{\alpha W}{f_{RR}}+\beta^rR'\,,\label{eq:evolve_R_GBSSN}
\end{align}
assuming that $f_{RR}\neq0$ holds. In fact, in the upcoming sections we shall be considering classes of $f(R)$ models such that $f_{RR}>0$ in order to avoid the so-called Dolgov-Kawasaki instability \cite{Dolgov:2003px}.

Finally, we note that the Hamiltonian constraint \eqref{eq:hamiltonian_constraint} in terms of the spherically symmetric variables is
\begin{align}\label{eq:hamiltonian_constraint_coordinatization}
    \mathcal{H} =& \chi^2\Bigg[\bar\Delta^r\phantom{}'+\frac{4}{r^2b}+\frac{2b'}{b^2r}-\frac{a''}{2a^2}+\frac{a'^2}{a^3}-\frac{4}{r^2a}+\frac{4\chi''}{\chi a}-\frac{6\chi'^2}{a\chi^2} \nonumber\\
    & -\frac{6b'}{rab}-\frac{b''}{ab}+\frac{4\chi'b'}{\chi ab}+\frac{8\chi'}{\chi ar} -\frac{2\chi'a'}{\chi a^2}-\frac{b'^2}{2ab^2}\Bigg]+\frac23K^2-\frac32A_a^2\nonumber\\
    & -R-\frac{1}{f_R}\Bigg[16\pi\rho-f+\frac{2\chi^2}{a}\left(f_R''-f_R'\left(\frac{a'}{2a}+\frac{\chi'}{\chi}-\frac{2}{r}-\frac{b'}{b}\right)\right)\Bigg]\,,
\end{align}
while the momentum constraint \eqref{eq:momentum_constraint} is
\begin{align}\label{eq:momentum_constraint_spherical_coordinates}
    \mathcal{M}_r = A_a'-\frac23K'+3A_a\left(\frac1r+\frac{b'}{2b}-\frac{\chi'}{\chi}\right)-\frac{1}{f_R}\left[8\pi j_r-W'-\left(A_a+\frac13K\right)f'_R\right]\,.
\end{align}

\subsection{Gauge conditions}
In the present work, we make us of the Bona-Mass\'o family of slicing conditions, i.e., we evolve the lapse function according to
\begin{align}\label{eq:evolution_lapse}
    \partial_t\alpha=-\alpha^2h(\alpha) K\,,
\end{align}
where $h(\alpha)$ is some function of the lapse \cite{Bona:1994dr,Baumgarte:2022auj}. More specifically, we will use either the $1+$log or harmonic slicing conditions, corresponding to $h(\alpha)=2/\alpha$ and $h(\alpha)=1$ respectively. For the shift vector, we introduce an auxiliary field $B^r$ and use the pair of equations
\begin{align}
    \partial_t\beta^r &= B^r\,,\label{eq:evolution_shift}\\
    \partial_t B^r &= \mu\Delta^r\,.\label{eq:evolution_B}
\end{align}
In equation \eqref{eq:evolution_B} $\mu$ is a constant parameter which we set either to $\mu=0$ or $\mu=3/4$ with the latter corresponding to the so-called Gamma-driver shift condition~\cite{Alcubierre:2002kk}.

\subsection{Klein-Gordon field}\label{sec:KG_decomposition}
For the specifc $f(R)$ models considered here, the non-GR related contributions vanish in vacuum space-times, and as such we need to impose a matter field to study any deviations from GR. As a simple model, we implement a massless KG field $\Phi=\Phi(x)$ described by the action
\begin{align}
    S_{\text m}=-\frac12\int\text{d}^4x\sqrt{-g}\ \left(\partial\Phi\right)^2\ \,,
\end{align}
whose well-known associated energy-momentum tensor and equations of motion are
\begin{align}
T_{\mu\nu}&=\partial_\mu\Phi\partial_\nu\Phi-\frac12g_{\mu\nu}\left(\partial\Phi\right)^2\,,\\
    \Box\Phi&=0\,,\label{eq:Phi_eom}
\end{align}
respectively. In terms of the spatial metric and unit normal vector field, the equation of motion \eqref{eq:Phi_eom} is
\begin{align}\label{eq:Phi_eom_spatial_metric}
    n^\mu n^\nu\nabla_\mu\nabla_\nu\Phi = \gamma^{\mu\nu}\nabla_\mu\nabla_\nu\Phi\ .
\end{align}
While we we will study the evolution of the GBSSN variables for the SKG system, the EKG system of equations has been studied numerically in spherical symmetry in~\cite{Werneck:2021kch,Choptuik:1992jv} using the ADM formalism, and in~\cite{alcubierre,Akbarian:2015oaa,Werneck:2021kch} using the GBSSN formalism. The EKG system has also been studied for the case of axial symmetry in~\cite{Healy:2013xia}.

Following the development in~\cite{alcubierre} for GR, we define the variables
\begin{align}\label{eq:psi_def}
    \Psi:=\Phi'\ ,
\end{align}
and
\begin{align}\label{eq:Pi_def}
    \Pi:=n^\mu\partial_\mu\Phi=\frac{1}{\alpha}\left(\partial_t\Phi-\beta^r\Psi\right)\ .
\end{align}
An evolution equation for $\Phi$ follows directly from the definition of $\Pi$, namely
\begin{align}\label{eq:evolve_Phi}
    \partial_t\Phi=\alpha\Pi+\beta^r\Psi\ .
\end{align}
A spatial derivative of this expression leads to an evolution equation for $\Psi$
\begin{align}\label{eq:evolve_Psi}
    \partial_t\Psi = \alpha'\Pi+\alpha\Pi'+\beta^r\phantom{}'\Psi+\beta^r\Psi'\ .
\end{align}
Then a Lie derivative of~\eqref{eq:Pi_def} determines the evolution equation of $\Pi$,
\begin{align}\label{eq:lie_Pi}
    \lie_{\bm n}\Pi& = n^\nu\nabla_\nu\left(n^\mu\nabla_\mu\Phi\right)\nonumber\\
    & = a^\mu D_\mu\Phi+n^\mu n^\nu\nabla_\mu\nabla_\nu\Phi\,,
\end{align}
where $\bm a$, as defined in equation \eqref{eq:deF_acceleration_vector}, is the acceleration vector. Substituting the equation of motion \eqref{eq:Phi_eom_spatial_metric} into \eqref{eq:lie_Pi} and relating the acceleration vector to the lapse function through $a_\mu=D_\mu\ln\alpha$ gives
\begin{align}
\lie_{\bm n}\Pi=D^\mu\ln\alpha D_\mu\Phi+D^2\Phi+K\Pi\ .
\end{align}

In terms of the metric components, this expression reads\footnote{We note that one could replace the first and second-order derivatives of $\Phi$ with $\Psi$ and its first order derivative respectively in the square bracket of equation \eqref{eq:evolve_Pi}. However, for our code using the PIRK scheme, we have found that using the first and second derivatives of $\Phi$ leads to improved numerical behaviour.}
\begin{align}\label{eq:evolve_Pi}
    \partial_t\Pi=\frac{\chi^2}{a}\alpha'&\Psi+\alpha K\Pi+\frac{\alpha\chi^2}{a}\bigg[\Phi''-\Phi' 
    \Big(\frac{a'}{2a}-\frac{b'}{b}-\frac{2}{r}+\frac{\chi'}{\chi}\Big)\bigg]+\beta^r\phantom{}\Pi'\ .
\end{align}
The matter variables enter into the $f(R)$ GBSSN evolution equations through the energy and momentum densities,
\begin{align}
    \rho=\frac12\left(\Pi^2+\frac{\chi^2\Psi^2}{a}\right)\ ,\qquad\text{and}\qquad
    j^r=-\frac{\chi^2\Psi\Pi}{a}\ ,
\end{align}
respectively as well as the stress-energy tensor with components
\begin{align}
S_r^{\ r}=\rho\ ,\ \ \ \ S_\theta^{\ \theta}=S_\psi^{\ \psi}=\frac12\left(\Pi^2-\frac{\chi^2\Psi^2}{a}\right)\ .
\end{align}
We also note that the $(r,r)$ component of the trace-free part of the stress-energy tensor, used in~\eqref{eq:evolution_A_a}, is given by
\begin{align}
S^{\text{TF}}_{rr}=\frac{2\Psi^2}{3}\ .
\end{align}

\section{Numerical results}\label{sec:numerical_simulations}

In this article, we perform numerical evolutions of three different scenarios: 
$(i)$ the evolution of the Schwarzschild black hole solution; $(ii)$ the evolution of flat (Minkowski) space-time endowed with non-trivial gauge dynamics; and finally $(iii)$ the evolution of a massless KG scalar field.
In the following, we shall consider these three scenarios in the context of the so-called \textit{Starobinsky $f(R)$ gravity models} for which the $f(R)$ function in the gravitational action \eqref{eq:gravitational_action} is of the form \cite{Vilenkin:1985md}
\begin{align}\label{eq:f_R_starobinsky}
f(R)=R+\frac{\ell}{2}R^2\ ,
\end{align}
where $\ell$ is a constant of dimension length squared\footnote{We note here that, for such a quadratic model, there will be non-trivial corrections to GR in the presence of matter. Referring the reader to the scalar-picture representation \eqref{eq:f_R_auxiliary_formulation} for $f(R)$ gravity, the field $\varphi$ will indeed be dynamical. However, we note that for the case of a vacuum, the Schwarzschild solution is the only static and spherically symmetric solution \cite{WHITT1984176}.
Also, it can be shown how in four dimensions
quadratic curvature invariants such as $R^{\mu\nu}R_{\mu\nu}$, $R^{\mu\nu\alpha\beta}R_{\mu\nu\alpha\beta}$ and $\varepsilon^{\mu\nu\sigma\gamma}R_{\mu\nu\alpha\beta}R^{\;\;\;\;\alpha\beta}_{\sigma\gamma}$ in a more general gravitational Lagrangian can be reduced to a term proportional to $R^2$. See for instance \cite{DeWitt:1964mxt} for further details.
}. For the cases $(i)$ and $(ii)$ above, there is no formal difference between the Starobinsky and pure GR scenarios since the evolution variables $R$ and $W$ are identically zero. These two tests, then, demonstrate the robustness of our implementation in the case of standard GR and can be compared with similar results from the literature~\cite{alcubierre,brown,Montero:2012yr,Baumgarte:2012xy}.

For the case of a massless KG scalar field, we anticipate that the non-GR contributions from the Starobinsky model will be relevant since the space-time Ricci scalar has non-zero initial data for such a case.

Regarding this third case, we note that as a result of the additional degrees of freedom from the fourth-order derivatives of the metric, the numerical implementations discussed in \cite{alcubierre,Montero:2012yr,Akbarian:2015oaa} are not immediately applicable. Firstly, and as will be discussed in Section \ref{sec:numerical_f_R}, the numerical methods employed in \cite{alcubierre} to solve the Hamiltonian constraint in the initial hypersurface are not suitable for the general $f(R)$ case. Secondly, it is necessary to introduce the evolution variables $R$ and $W$ into the numerical scheme according to the GBSSN equations presented in Section \ref{sec:reduction_to_spherical_symmetry}.

\subsection{Schwarzschild solution}\label{sec:schwarzschild_wh}

A standard test case for spherically symmetric evolution codes is the  Schwarzschild solution which exercises the bulk of the evolution equations with non-trivial gauge dynamics and presents some challenges in regularizing the origin. As initial data, we impose a conformally flat 3-metric by setting $a(0,r)=b(0,r)=1$, zero extrinsic curvature $K(0,r)=A_a(0,r)=0$ and also set $\bar\Delta^r(0,r)=0$. For the conformal factor $\chi$ we set 
\begin{align}
\chi(0,r)=\left(1+\frac{M}{2r}\right)^{-2}\,,
\end{align}
where $M$ is the mass of the black hole and has dimensions of length. Such initial data corresponds to the Schwarzschild solution in isotropic coordinates which possesses a wormhole topology. The evolution of such a configuration has been used as a test-bed in a number of studies~\cite{alcubierre,brown,Montero:2012yr}. The initial lapse is set to $\alpha(0,r)=\chi(0,r)$ and evolved using the $1+\log$ slicing condition, corresponding to equation~\eqref{eq:evolution_lapse} with $h(\alpha)=2/\alpha$. The shift vector is initially set to zero, i.e., $\beta^r(0,r)=0$, and evolved using the Gamma-driver condition~\eqref{eq:evolution_shift}--\eqref{eq:evolution_B} with $\mu=3/4$.

The spatial grid straddles the origin with gridpoints located at $r_n=\left(n-1/2\right)\Delta r$
with a base resolution of $\Delta r=0.0125M$ and time-step of $\Delta t=\Delta r/2$. We use even boundary conditions at the origin for scalar functions and components of type $(0,2)$ tensor fields and odd boundary conditions for vectors. We evolve variables using the RK4 scheme. Spatial derivatives are calculated using second-order finite differences; using forward differences for the shift advection terms and central differences for all others. At the outer boundary, we freeze the data associated with the last two grid points; similar to what is done in \cite{brown}.

\begin{figure}[!htb]\centering
    \begin{subfigure}[b]{0.49\textwidth}
        \includegraphics[width=\textwidth]{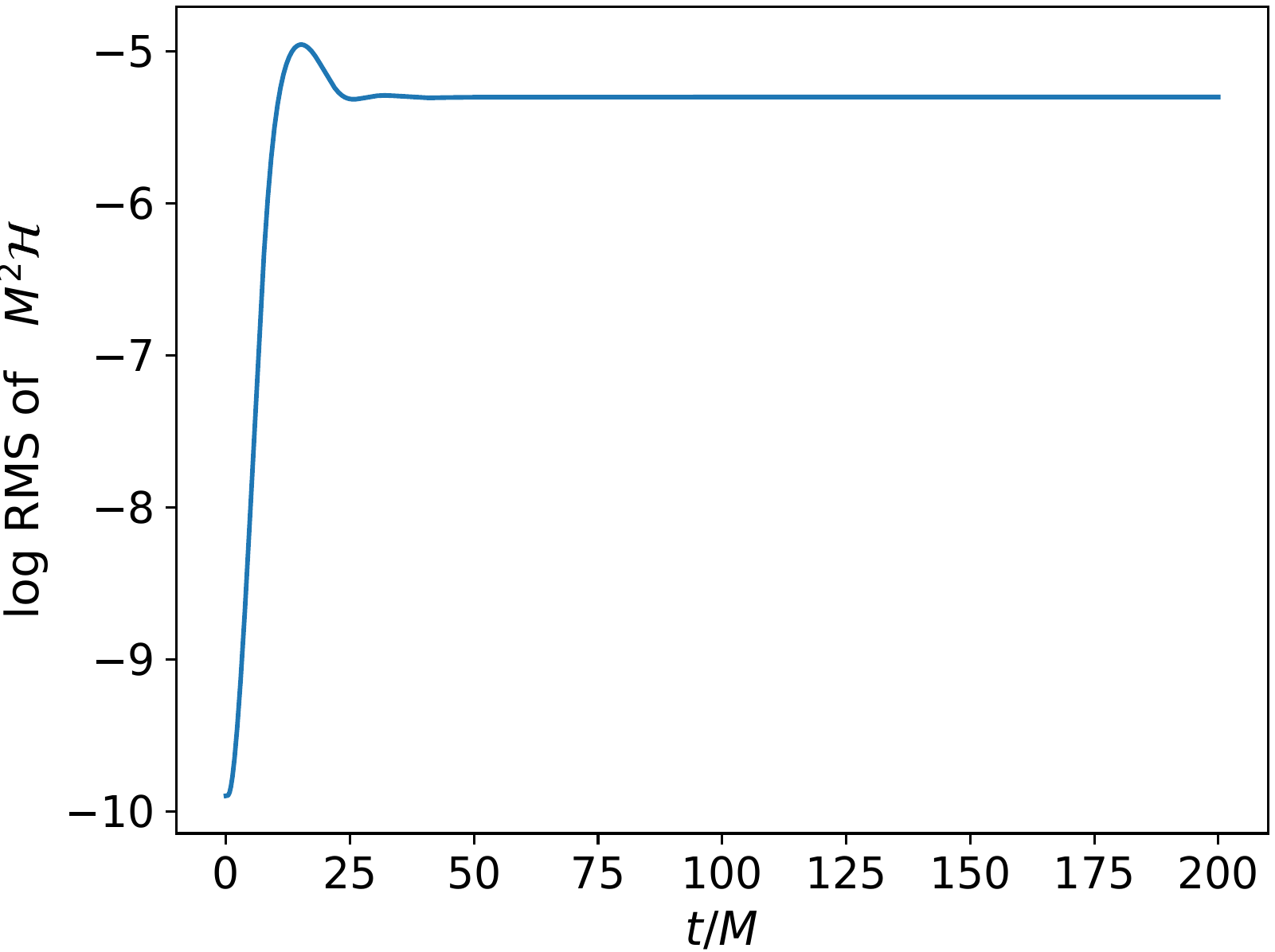}
    \end{subfigure}
    \begin{subfigure}[b]{0.49\textwidth}
        \includegraphics[width=\textwidth]{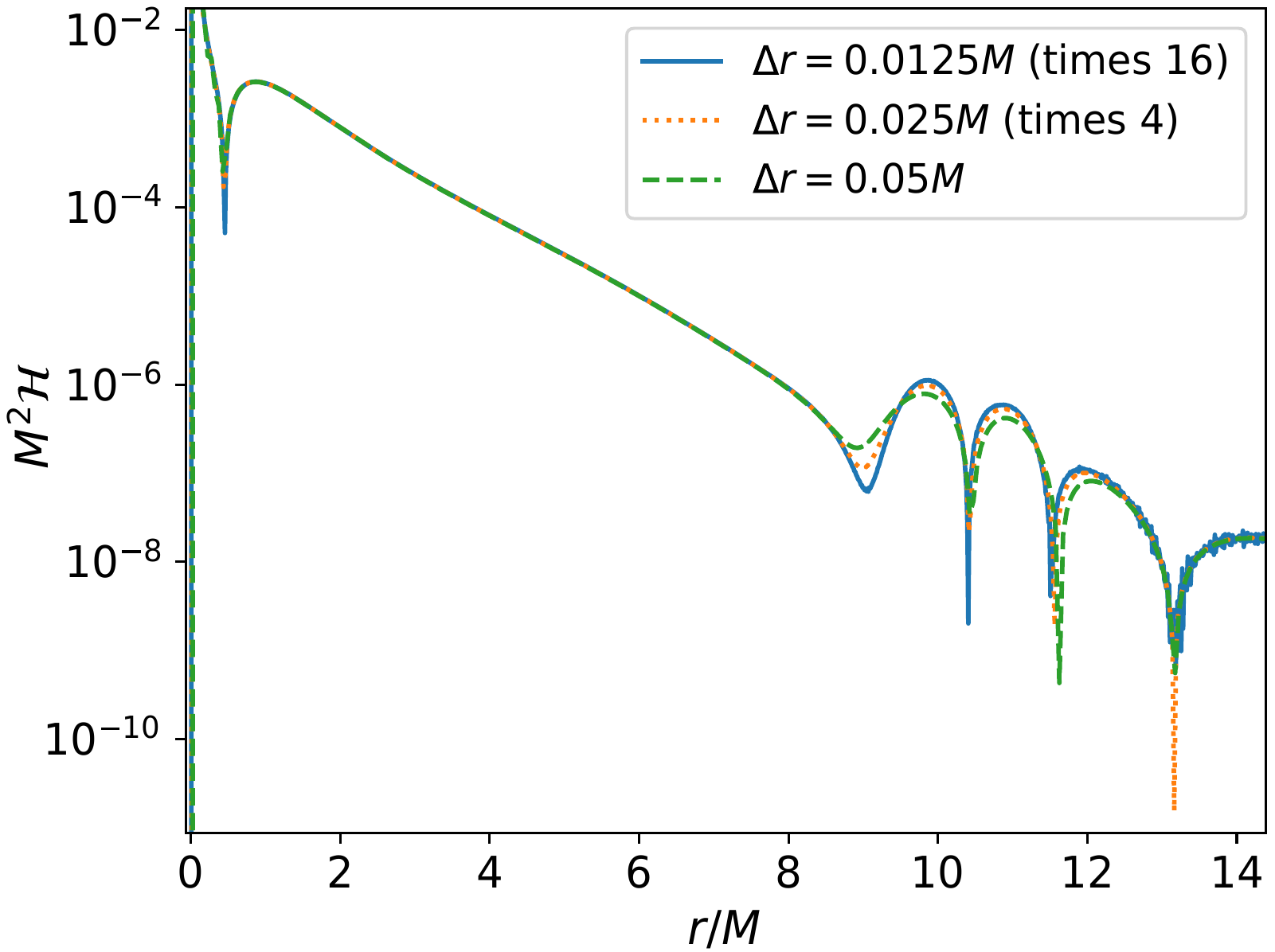}
    \end{subfigure}
    \caption{Hamiltonian constraint \eqref{eq:hamiltonian_constraint_coordinatization} for the evolution of Schwarzschild wormhole initial data. The left panel shows the natural logarithm of the RMS of the Hamiltonian constraint when we have used a spatial resolution of $\Delta r=0.0125M$ and a time-step of $\Delta t=\Delta r/2$. In the right panel, we have plotted the spatial profiles of the Hamiltonian constraint at $t=10M$. The solid blue, dotted orange and dashed green curves show the Hamiltonian constraint for spatial resolutions of $\Delta r=0.0125M$ (rescaled by a factor of $16$), $0.025M$ (rescaled by a factor of 4) and $0.05M$ respectively. The rescalings used are consistent with second-order convergence.}
    \label{fig:ham-bh}
\end{figure}

In the left panel of Figure~\ref{fig:ham-bh} we plot the natural logarithm of the root-mean-square (RMS) of the Hamiltonian constraint \eqref{eq:hamiltonian_constraint_coordinatization} over the entire numerical grid as a function of time, noting that it stabilizes at a value of less than $10^{-5}$ at the base resolution. A spatial profile at $t=10M$ is shown to the right, and is evaluated at three different resolutions to demonstrate second-order convergence of the error. The solid blue, dotted orange and dashed green curves correspond to spatial resolutions of $\Delta r=0.0125M$, $0.025M$ and $0.05M$ respectively. The profiles corresponding to spatial resolutions of $\Delta r=0.0125M$ and $\Delta r=0.025M$ have been rescaled by factors of $16$ and $4$ respectively; consistent with second-order convergence. We note that for each spatial resolution used, we have kept the time-step as $\Delta t=\Delta r/2$.

\begin{figure}[!htb]\centering
    \includegraphics[width=0.8\linewidth]{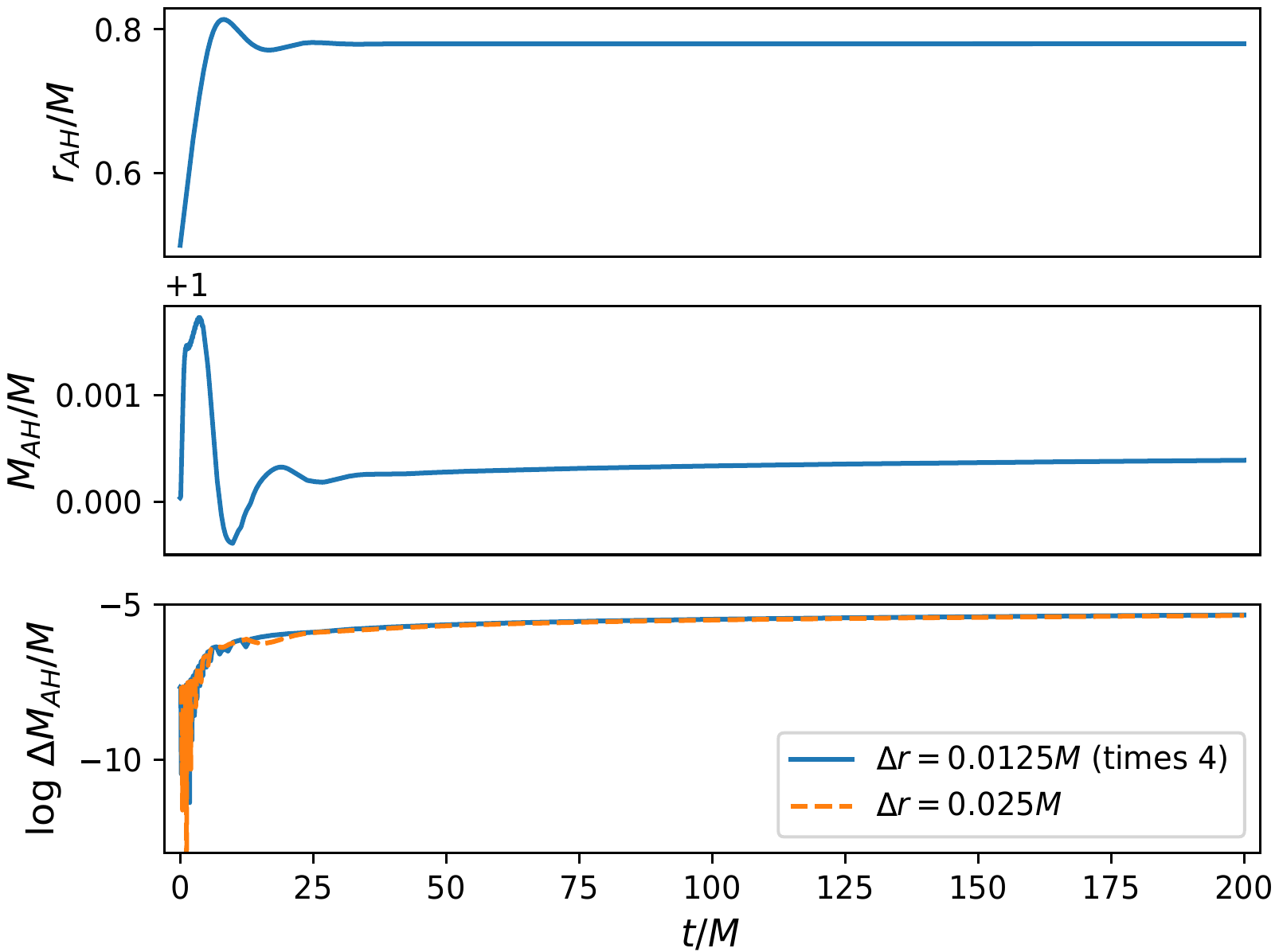}
    \caption{For the evolution of Schwarzschild initial data, the horizon is an effective diagnostic. In the top panel we see that the horizon radius grows from its initial value of $r=M/2$ due to gauge dynamics, but within 25$M$ has settled to an effectively constant position. The horizon mass shown in the centre panel similarly experiences a transient oscillation, though remains within $0.2\%$ of $M$ throughout the evolution and within $0.04\%$ at the final time of $t=200M$. In the bottom panel, we plot the difference $\Delta M_{\text{AH}}:=M_{\text{AH},2\Delta r}-M_{\text{AH},\Delta r}$ where $M_{\text{AH},2\Delta r}$ and $M_{\text{AH},\Delta r}$ denote the horizon mass computed using a spatial resolution of $\Delta r$ and $2\Delta r$ respectively. The solid blue curve, which is rescaled by a factor of 4, and the dashed orange curve show this difference when the spatial resolution is $\Delta r=0.0125M$ and $\Delta r=0.025M$ respectively. This bottom panel shows that the errors in the mass estimate converge away at second-order.}
    \label{fig:horizon-bh}
\end{figure}

Aspects of the apparent horizon evolution are plotted in Figure \ref{fig:horizon-bh}, including the apparent horizon position and mass evolution in the first and second panels respectively. The horizon radius $r_{\text{AH}}$ is computed by evaluating the expansion of outgoing null rays
\begin{align}
\Theta(r):=\frac{\chi}{\sqrt a}\left(\frac{2}{r}+\frac{b'}{b}-\frac{2\chi'}{\chi}\right)+A_a-\frac{2K}{3}\ ,
\end{align}
and numerically solving for $\Theta(r_{\text{AH}})=0$~\cite{thornburg}. Once the apparent horizon position has been obtained, we compute the mass using
\begin{align}
M_{\text{AH}}=\sqrt{\frac{A_{\text{AH}}}{16\pi}}=\frac{\sqrt{r^2b\sqrt{a}}}{2\chi}\bigg|_{r=r_{\text{AH}}}\,,
\end{align}
where $A_{\text{AH}}$ is the area of the apparent horizon. The deviation of the mass from its initial value is small; remaining within $0.2\%$ of $M$ throughout the simulation and within $0.04\%$ at $t=200M$. In the third panel, we show the convergence of the apparent horizon mass. Therein, we have plotted the natural logarithm of the difference, $\Delta M_{\text{AH}}:=M_{\text{AH},2\Delta r}-M_{\text{AH},\Delta r}$ where $M_{\text{AH},2\Delta r}$ is the apparent horizon mass computed using a spatial resolution of $2\Delta r$ and $M_{\text{AH},\Delta r}$ is the mass computed using a spatial resolution of $\Delta r$ and interpolated onto a grid with spacing $2\Delta r$. The solid blue curve shows this difference rescaled by a factor of $4$ for the case where $\Delta r=0.0125M$ while the dashed orange curve shows this difference when $\Delta r=0.025M$.

\subsection{Gauge pulse in flat space}\label{sec:gauge_pulse}

Non-trivial gauge dynamics in the evolution of flat (Minkowski) space-time can be induced in the metric variables by using the following initial data for the lapse function~\cite{alcubierre}
\begin{align}\label{eq:lapse_gauge_pulse}
\alpha(0,r)=1+\frac{0.01r^2}{1+r^2}\left[{\text e}^{-(r-d)^2/\sigma^2}+{\text e}^{-(r+d)^2/\sigma^2}\right]\,,
\end{align}
where $d$ and $\sigma$ have dimensions of length and are the position and width of the initial profile respectively. For our consideration, we take $d/\sigma=5$. By inserting such initial data into the numerical code described in the previous subsection, the GBSSN variables $K$, $A_a$ and $\bar\Delta^r$ start to diverge early on and we end up with NaN (not a number) values; indicating the presence of instability. We note that we have experimented with both 1+log slicing and harmonic slicing for the lapse function while using either a zero shift vector or the Gamma-driver condition and, regardless of the combination of these gauge choices, the aforesaid instability is noticed\footnote{While the instability occurs for any of the combination of gauge choices mentioned above, we notice that $K$ diverges slightly faster when using harmonic slicing and $\bar\Delta^r$ diverges faster when using the Gamma-driver condition.}. The unsuitability of the numerical scheme outlined in the previous subsection to evolve a gauge pulse in flat space using spherical polar coordinates has been noted in \cite{alcubierre}. To perform the stable evolution of a gauge pulse, a number of approaches are possible \cite{alcubierre,Montero:2012yr}. One such approach would be to perform a regularization by introducing new evolution variables through a redefinition of fields \cite{alcubierre,Werneck:2021kch}. Another approach involves the application of some partially implicit numerical scheme as opposed to the RK4 method \cite{Montero:2012yr,Akbarian:2015oaa}. In the present work, we use the latter approach by implementing the PIRK method used in \cite{Montero:2012yr}.

Thus, by following \cite{Montero:2012yr}, we evolve $a,b,\alpha,\beta^r$ and $\chi$ explicitly. Subsequently, we then evolve the extrinsic curvature variables $K$ and $A_a$ partially implicitly using updated values of the conformal 3-metric components. We then evolve $\bar\Delta^r$ using updated values of the conformal 3-metric components and the extrinsic curvature, and finally evolve the auxiliary field $B^r$. For a detailed outline of the PIRK scheme used here we direct the reader to~\ref{sec:appendix_PIRK_gauge}. Although this procedure allows for non-vanishing shift conditions, as in the Gamma-driver condition, we set the shift to zero for our gauge pulse evolutions. We also note that we make use of $1+\text{log}$ slicing for our gauge pulse evolutions.

\begin{figure}[!htb]\centering
\includegraphics[width=0.8\linewidth]{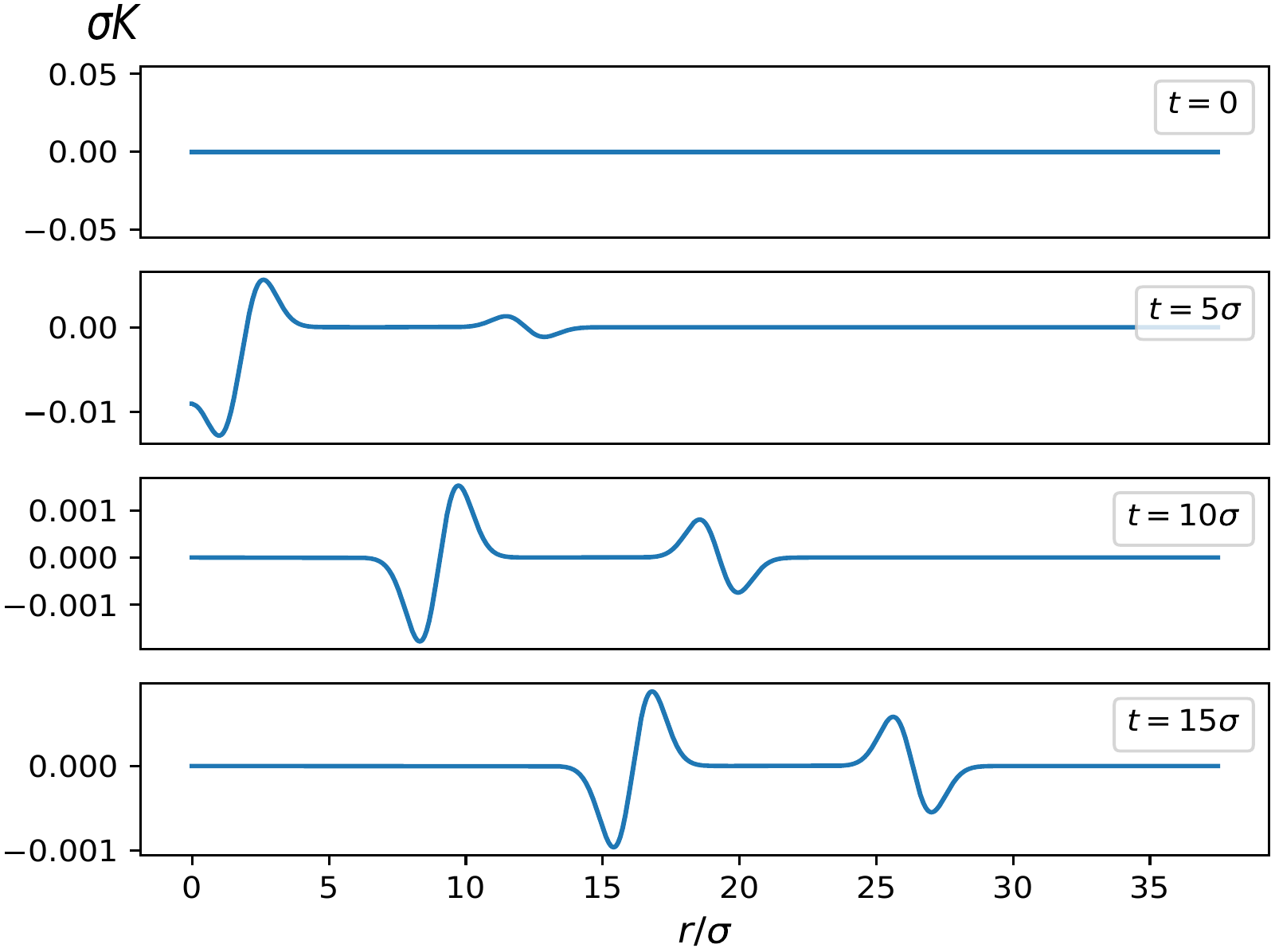}
\caption{Results corresponding to the evolution of a gauge pulse \eqref{eq:lapse_gauge_pulse} in the lapse function in flat space using $1+\text{log}$ slicing with a vanishing shift vector. Here, we plot the evolution of the trace, $K$, of the extrinsic curvature tensor. The first, second, third and fourth panels show the spatial profiles of $K$ at $t=0$, $5\sigma$, $10\sigma$ and $15\sigma$ respectively. To produce these plots, we have used a space-step of $\Delta r=0.0125\sigma$ and a time-step of $\Delta t=\Delta r/2$.}
\label{fig:K_flat}
\end{figure}

\begin{figure}[!htb]\centering
    \begin{subfigure}[b]{0.49\textwidth}
        \includegraphics[width=\textwidth]{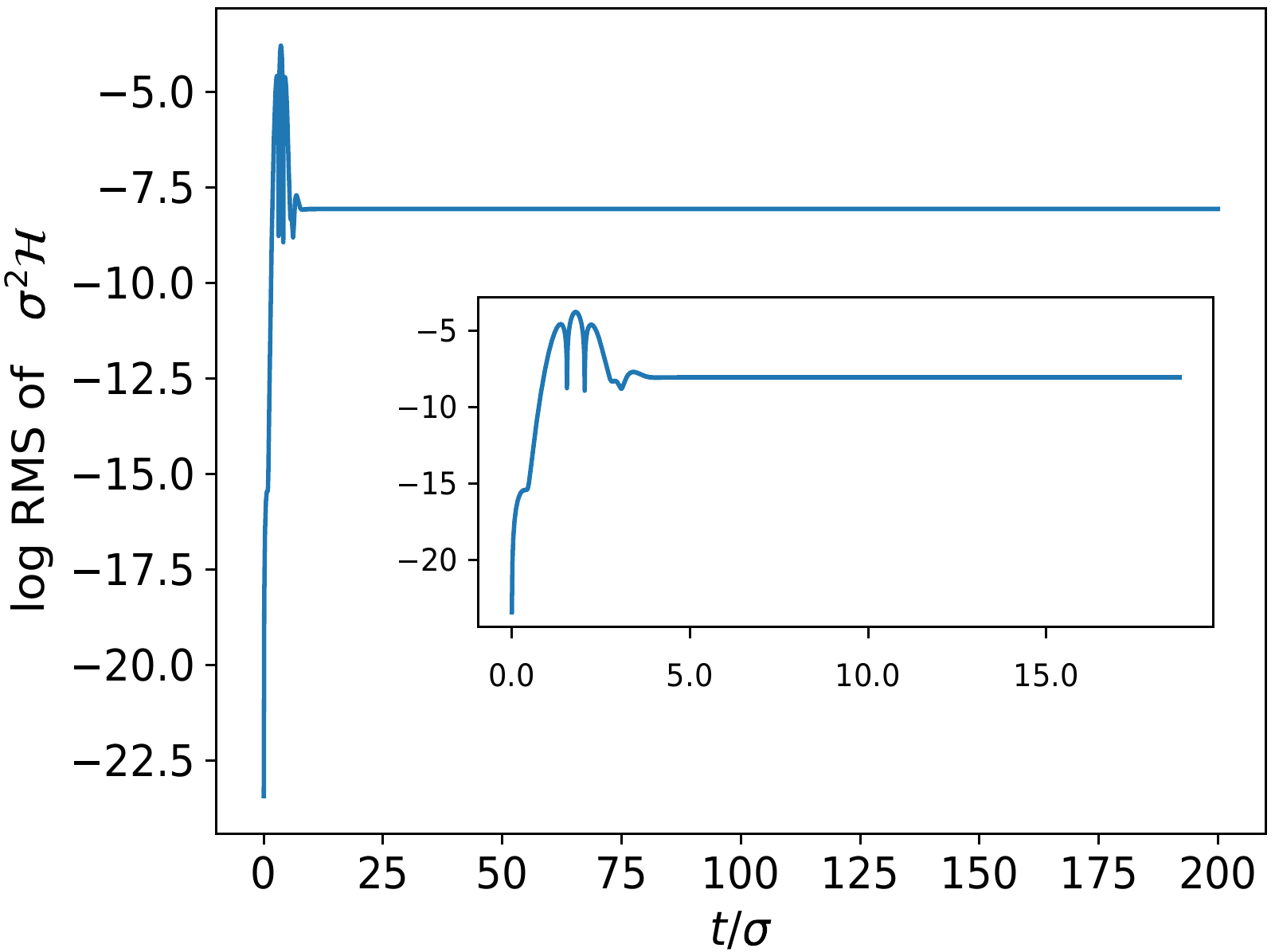}
    \end{subfigure}
    \begin{subfigure}[b]{0.49\textwidth}
        \includegraphics[width=\textwidth]{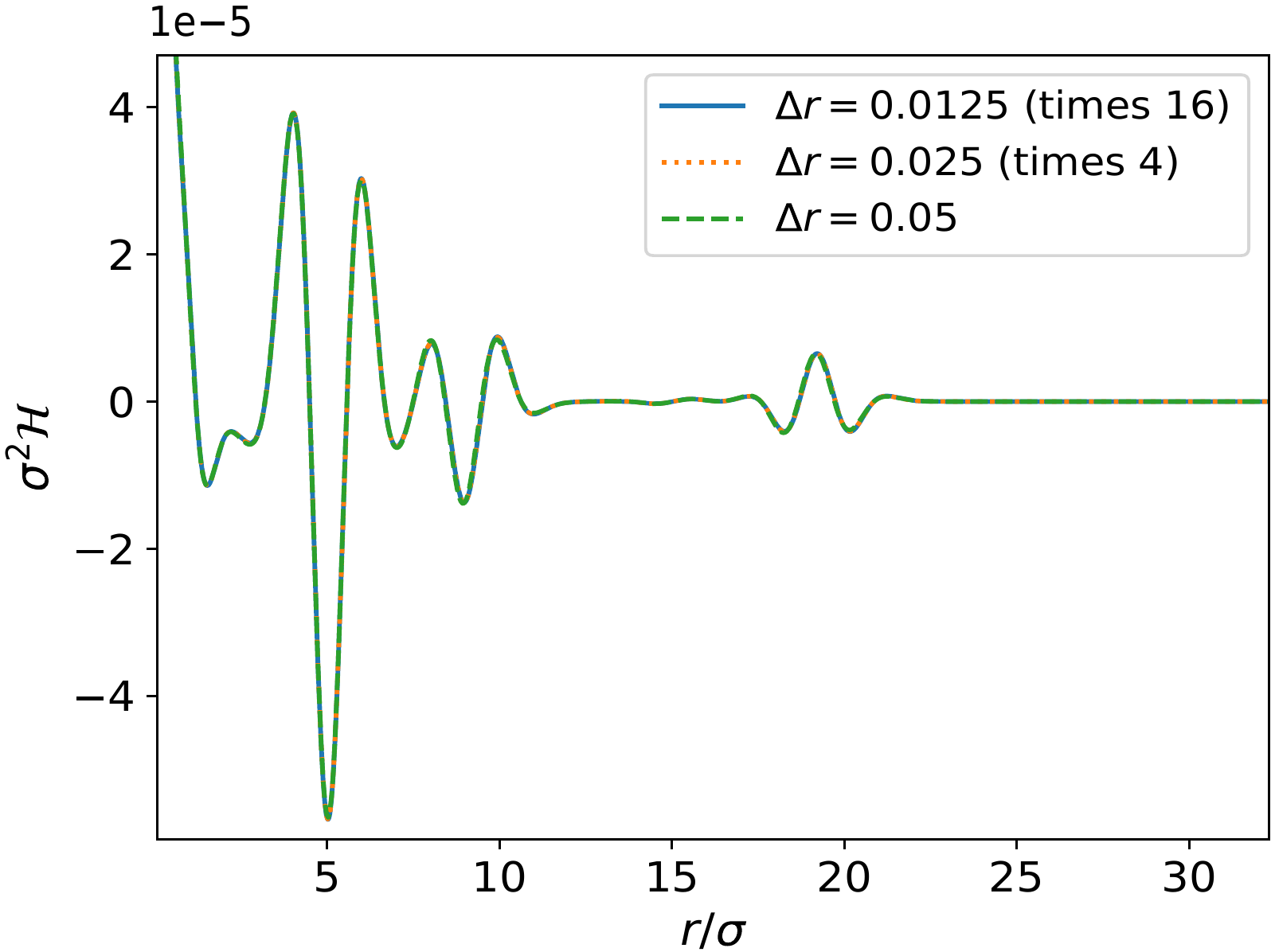}
    \end{subfigure}
\caption{Results corresponding to the evolution of a gauge pulse \eqref{eq:lapse_gauge_pulse} in the lapse function in flat space using $1+\text{log}$ slicing with a vanishing shift vector. Here, we have plotted the spatial profile of the Hamiltonian constraint at $t=10\sigma$ for three different spatial resolutions: $\Delta r=0.0125\sigma$, $0.025\sigma$ and $0.05\sigma$. The profiles corresponding to space-steps of $\Delta r=0.0125\sigma$ and $0.025\sigma$ have been rescaled by factors of $16$ and $4$ respectively; consistent with second order convergence. To obtain these results, we have used a time-step of $\Delta t=\Delta r/2$ for each choice of spatial resolution. The fact that the lines in this figure lie on top of each other demonstrates the second-order convergence of our numerical relativity code for the case of the evolution of a gauge pulse in flat space. }
\label{fig:ham_flat}
\end{figure}

In Figures \ref{fig:K_flat} and \ref{fig:ham_flat} we give the results of the evolutions for the case of flat space-time endowed with non-trivial gauge dynamics. Figure \ref{fig:K_flat} shows spatial profiles of the extrinsic curvature trace $K$ at times $t=0$, $5\sigma$, $10\sigma$ and $15\sigma$; produced using a space-step of $\Delta r=0.0125\sigma$ and a time-step of $\Delta t=\Delta r/2$. In Figure \ref{fig:ham_flat}, we plot the natural logarithm of the RMS of the Hamiltonian constraint, in the left panel, up to $t=200\sigma$; demonstrating the long-term stability of our code for this case. In the right panel, we show the spatial profiles of the Hamiltonian constraint for three resolutions. The solid blue, dotted orange and dashed green curves are produced using spatial resolutions of $\Delta r=0.0125\sigma$, $\Delta r=0.025\sigma$ and $\Delta r=0.05\sigma$ respectively while keeping $\Delta t=\Delta r/2$ for each case. For the cases where $\Delta r=0.0125\sigma$ and $0.025\sigma$, we rescale the profiles by $16$ and $4$ respectively which correspond to second-order convergence. It is evident from Figure \ref{fig:ham_flat} that these profiles are almost indistinguishable; indicating the second-order convergence of our numerical code for the case of flat space non-trivial gauge dynamics.

\subsection{Scalar field evolution}\label{sec:numerical_f_R}
Having verified that our numerical code correctly simulates the evolution of both the Schwarzschild solution as well as non-trivial gauge dynamics in flat space, we now turn our attention to the evolution of a massless KG scalar field in a dynamical space-time as generated by the quadratic $f(R)$ Starobinsky model. While the space-time Ricci scalar had initial data of zero for the two previous cases, $R$ will be non-zero for the case of a massless KG field. As a result, there will be non-trivial effects arising from the non-GR contribution of the $f(R)$ Starobinsky model.

Indeed, we set the initial data for the KG scalar field using
\begin{align}\label{eq:sf_initial_data}
\Phi(0,r)=p\ {\text e}^{-\left(r-d\right)^2/\sigma^2}\ ,\qquad \Pi(0,r)=0\,,
\end{align}
where $p$ is the dimensionless initial amplitude of the Gaussian profile while $d$ and $\sigma$ have dimensions of length and give, respectively, the position and width of the profile. The initial data for the variable $\Psi$ is computed analytically by taking the spatial derivative of $\Phi(0,r)$. For all of the subsequent numerical considerations, we set the initial position at $d/\sigma=5$.

We assume a conformally flat initial geometry, setting $a=b=1$ and $\bar\Delta^r=\beta^r=K=A_a=W=0$. In the initial slice, the space-time Ricci scalar and the three-dimensional Ricci scalar coincide, i.e., we have $R(0,r)=\!^3R(0,r)$. In terms of the conformal factor, the initial Ricci scalar yields
\begin{align}\label{eq:varphi_to_Ricci}
    R(0,r)=4\chi\left(\chi''-\frac{3\chi'\phantom{}^2}{2\chi}+\frac{2\chi'}{r}\right)\ .
\end{align}

Initial data for $\chi$, and hence for $R$, requires that we solve the Hamiltonian constraint with the given initial matter profile \eqref{eq:sf_initial_data}. We note that the momentum constraint~\eqref{eq:momentum_constraint_spherical_coordinates} is identically zero in the initial slice when substituting in the aforementioned initial data for the extrinsic curvature, matter field and the new field $W$. From the $f(R)$ Hamiltonian constraint equation \eqref{eq:hamiltonian_constraint_coordinatization} we have
\begin{align}\label{eq:hamiltonian_id}
    f\left(R\right)-2\chi^2\left[f_R''\left(R\right)-f_R'\left(R\right)\left(\frac{\chi'}{\chi}-\frac2r\right)\right]=16\pi\rho\ ,
\end{align}
in the initial slice. For the Starobinsky gravity model \eqref{eq:f_R_starobinsky} equation \eqref{eq:hamiltonian_id} gives us
\begin{align}\label{eq:hammy_solve}
R+\frac{\ell}{2}R^2-2\ell\chi^2\bigg[R''-R'\left(\frac{\chi'}{\chi}-\frac2r\right)\bigg]-8\pi\chi^2\Psi^2=0\ ,
\end{align}
where the Ricci scalar is related to the conformal factor through equation \eqref{eq:varphi_to_Ricci}. In the case of GR ($\ell=0$), and as mentioned in \cite{alcubierre}, one can perform a redefinition of fields based on $1/\sqrt\chi$ so that equation \eqref{eq:hammy_solve} is a linear differential equation in the aforesaid redefined variable.
For the Starobinsky model such that $\ell\neq0$, this resulting equation is, however, non-linear. Here we have used a Newton's method approach to iteratively solve \eqref{eq:hammy_solve} for the conformal factor; using second-order finite differencing for the derivatives. Upon obtaining the initial conformal factor, the initial Ricci scalar is determined by \eqref{eq:varphi_to_Ricci}.

We now turn our attention to the evolution of this initial data. We once again take the shift vector to be zero throughout the simulation while evolving the lapse function using the harmonic slicing condition, i.e., we use equation \eqref{eq:evolution_lapse} with $h(\alpha)=1$. In the case of $\ell=0$, i.e., for GR, we evolve the standard GBSSN variables consisting of the metric components, the trace and trace-free part of the extrinsic curvature tensor, the regularized conformal connection function and the three matter fields. To evolve this system using the PIRK method, we need to introduce the matter variables $\Phi$, $\Psi$ and $\Pi$ into the scheme. Here, still in the pure GR scenario, we follow the procedure of treating the fields $\Phi$ and $\Psi$ explicitly, and therefore evolve these in the same way as the metric components, and evolve the field $\Pi$ partially implicitly after the extrinsic curvature terms and before the regularized conformal connection function. For a detailed discussion on how the matter fields are evolved using the PIRK scheme, the reader is directed to \ref{sec:PIRK_matter_detailed}. 

In order to evolve the SKG system numerically, we need to evolve the fields $R$ and $W$ within the PIRK scheme. It is now a question of whether we treat these fields explicitly or partially implicitly. We have found that, when treating both $R$ and $W$ explicitly, i.e., updating these along with the metric components, and evolving the SKG system, the variable $R$ quickly diverges followed by the divergence of $A_a$, $K$, $\bar\Delta^r$ and then the remaining evolution variables; resulting in unstable evolution. However, when updating $R$ explicitly and updating $W$ partially implicitly along with the extrinsic curvature variables, we obtain the stable evolution of the SKG system. In \ref{prik_f_R_appendix} we discuss in detail how we introduce the variables $R$ and $W$ into the PIRK scheme for the non-trivial $f(R)$ scenario.

We now turn our attention to demonstrating the stability and convergence of our numerical code for the case of a massless KG scalar field and a non-zero value for $\ell$ in \eqref{eq:f_R_starobinsky}. As a test case, we specify the Starobinsky gravity model by setting $\ell/\sigma^2=10^{-4}$ and specify the initial data for the matter by setting $p=0.01$ in \eqref{eq:sf_initial_data}. We consider two different spatial resolutions: $\Delta r=0.025\sigma$ and $\Delta r=0.05\sigma$ while keeping $\Delta t=\Delta r/2$ and track the RMS of the Hamiltonian constraint. After obtaining the initial data for the conformal factor when $\Delta r=0.025\sigma$, we subsequently interpolate the obtained profile to find the initial data for the geometry when $\Delta r=0.05\sigma$. In doing this, we ensure that the initial data is the same for both spatial resolutions.

\begin{figure}
    \centering
    \includegraphics[width=0.6\textwidth]{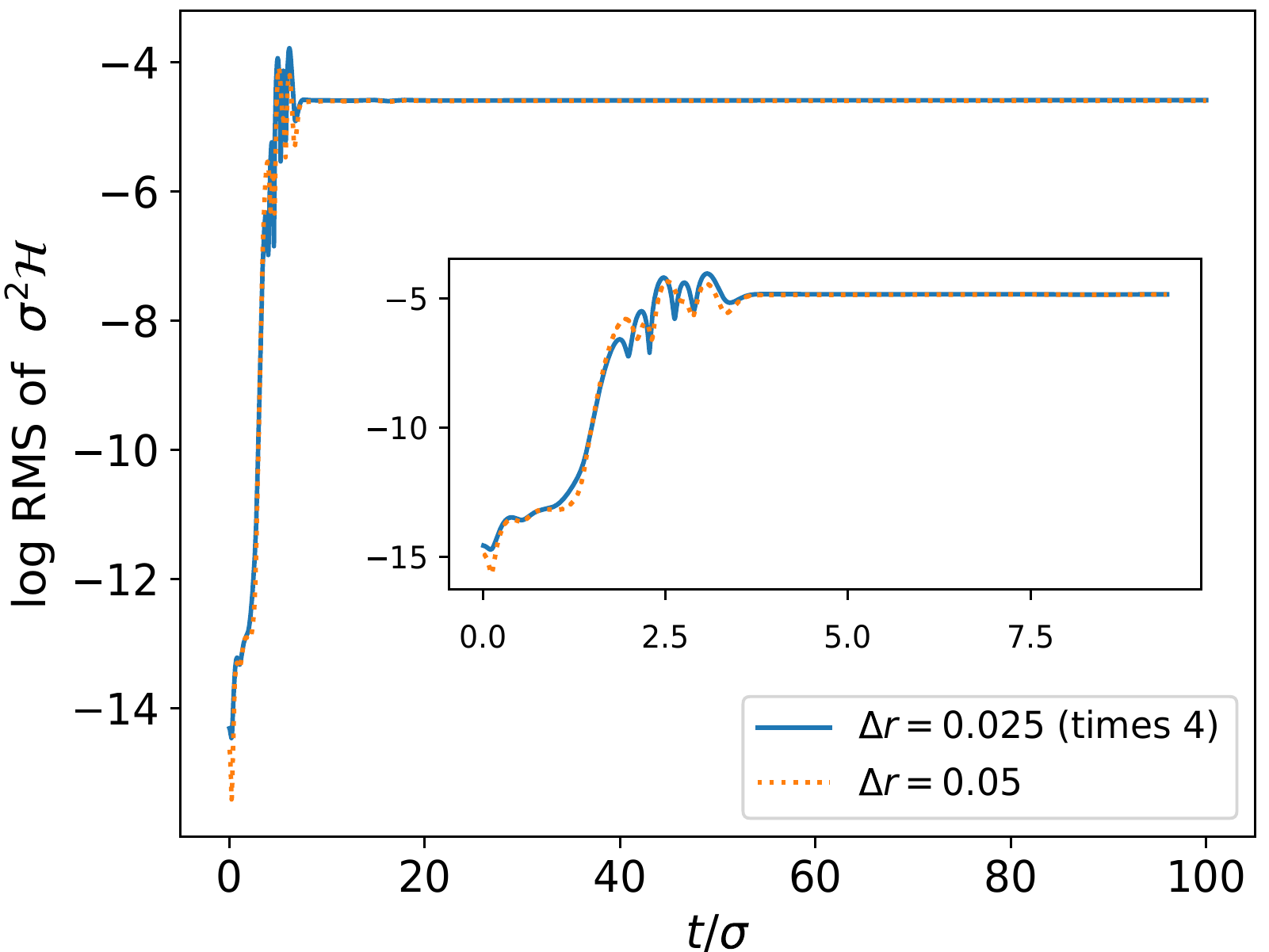}
    \caption{Natural logarithm of the RMS of the Hamiltonian constraint for the evolution of a massless KG scalar field in a dynamical space-time generated by the quadratic Starobinsky $f(R)$ model with $\ell/\sigma^2=10^{-4}$ in \eqref{eq:f_R_starobinsky}. For the initial data, we have set $p=0.01$ in \eqref{eq:sf_initial_data}. We have plotted the RMS of the Hamiltonian constraint using space-steps of $\Delta r=0.025\sigma$ and $\Delta r=0.05\sigma$; rescaling the former by a factor of 4 which is consistent with second-order convergence.}
    \label{fig:ham_f_R}
\end{figure}

In Figure \ref{fig:ham_f_R}, we plot the natural logarithm of the RMS of the Hamiltonian constraint for the two specified spatial resolutions; rescaling the RMS associated with the higher resolution by a factor of 4 which corresponds to second-order convergence. In computing the RMS of the Hamiltonian constraint, we ignore the first two points next to the origin for the $\Delta r=0.025\sigma$ case while ignoring the first grid point next to the origin for the $\Delta r=0.05\sigma$ case. Therefore, the results of Figure \ref{fig:ham_f_R} indicate that the numerical code is second-order convergent for $r/\sigma>0.05$. This further indicates that there is a significant amount of numerical error contained very close to the origin as a result of the $1/r$ coordinate singularity; similar to what occurs in the vacuum cases of the previous subsections. We also note that we have evolved the system to $t=100\sigma$ to demonstrate the stability of our code.

Next, we compare the evolutions for a fixed $p$ and different values for $\ell$, i.e., we use the same initial data for the matter and consider different realizations within the class of Starobinsky $f(R)$ models. We expect that the most notable differences in the evolutions will occur for high curvatures in subcritical evolutions, i.e., no apparent horizon is detected. We set $p=0.04$ and carry out evolutions for the GR case ($\ell=0$) as well as for a number of $\ell\neq0$ cases. 
It is only for the non-GR ($\ell\neq0$) cases that $R$ and $W$ are used as evolution variables since the numerical scheme requires the condition $f_{RR}\neq0$ to be satisfied and where in fact we have assumed $f_{RR}>0$ in order to avoid the so-called Dolgov-Kawasaki instability being present as mentioned above. For the GR ($\ell=0$) case we evolve the standard set of GBSSN evolution variables that excludes $R$ and $W$. For these evolutions we use a spatial resolution of $\Delta r=0.01\sigma$ and a time-step of $\Delta t=\Delta r/2$.

\begin{figure*}[!htb]
\minipage{0.5\textwidth}
  \includegraphics[width=\linewidth]{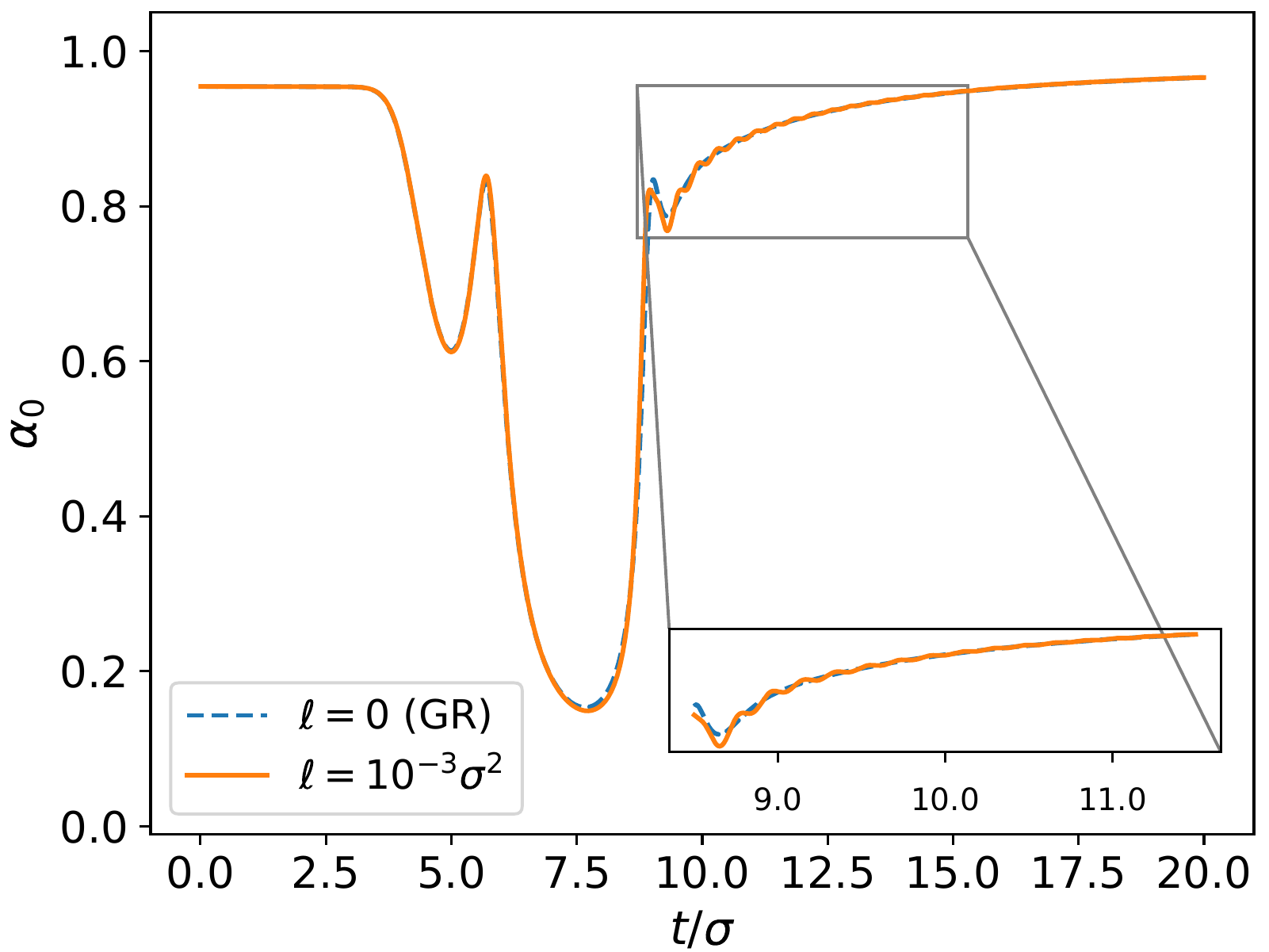}
\endminipage\hfill
\minipage{0.5\textwidth}
  \includegraphics[width=\linewidth]{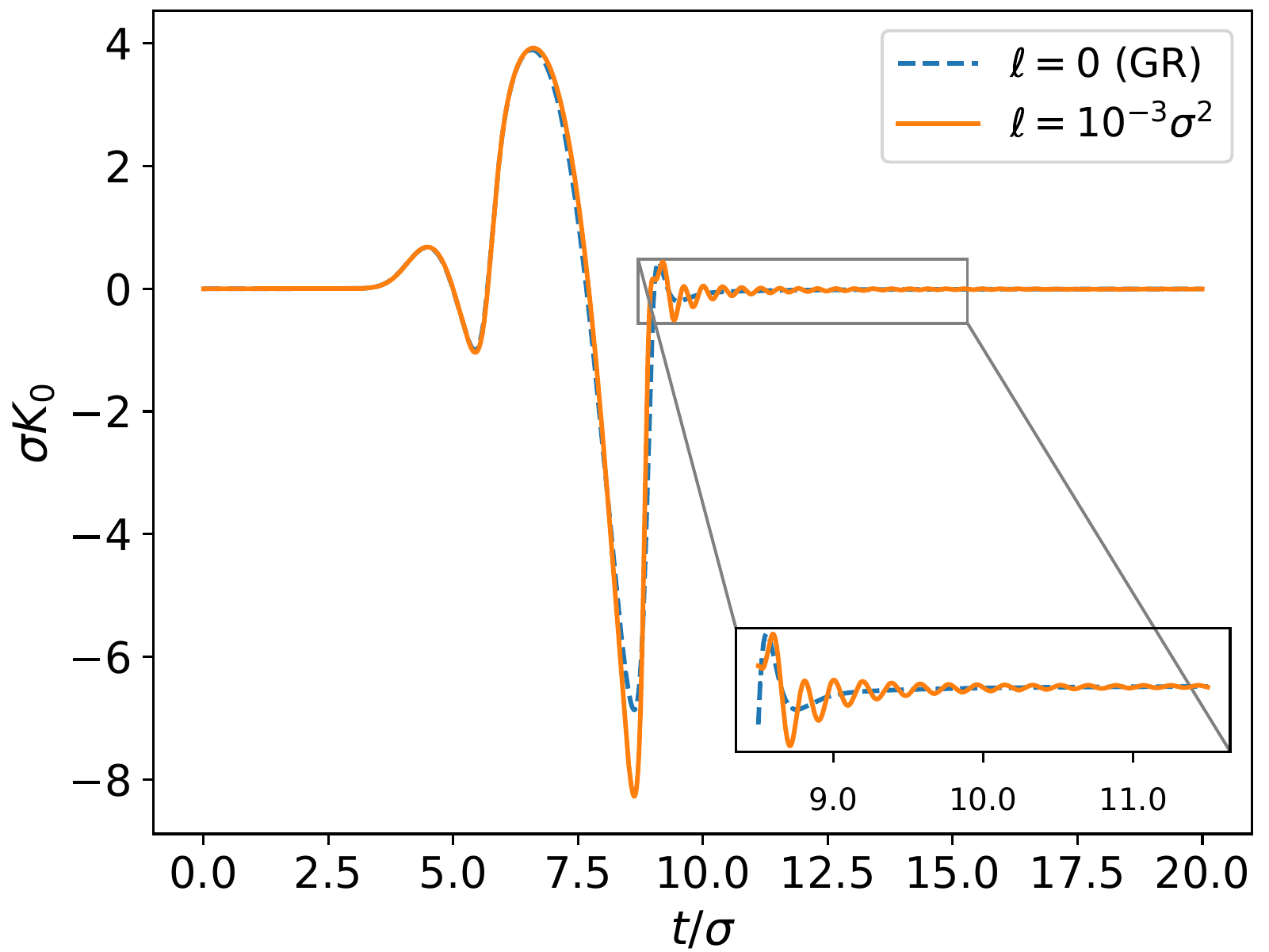}
\endminipage\hfill
\caption{Central values of the lapse function $\alpha_0$ and the trace of the extrinsic curvature tensor $K_0$ for the $f(R)$ parameter values $\ell=0$ and $\ell/\sigma^2=10^{-3}$ and we have set $p=0.04$. To produce these plots, we have used a space-step of $\Delta r=0.01\sigma$ and a time-step of $\Delta t=\Delta r/2$. The dashed blue curves are associated with the case where $\ell=0$ while the solid orange curves are associated with that of $\ell/\sigma^2=10^{-3}$. In both figures, we see that the extrema are greater in magnitude when $\ell/\sigma^2=10^{-3}$. In addition, around $t\sim10\sigma$, we notice the occurrence of damped oscillations for the case where $\ell/\sigma^2=10^{-3}$.}
\label{fig:lapse_f_R}
\end{figure*}

In Figure \ref{fig:lapse_f_R} we plot the central value of the lapse function $\alpha_0$ and the trace of the extrinsic curvature tensor $K_0$ over time for the cases of $\ell=0$ and $\ell/\sigma^2=10^{-3}$. The dashed blue curves correspond to the case where $\ell=0$ (GR) and the solid orange curves correspond to the case where $\ell/\sigma^2=10^{-3}$. The profiles are almost identical for the early portion of the evolution. However, the extrema, which occur at around $t\sim8\sigma$ and $t\sim9\sigma$, are greater in magnitude for the case where $\ell/\sigma^2=10^{-3}$. Also for the case of $\ell/\sigma^2=10^{-3}$, we note the appearance of damped oscillations beginning at around $t\sim10\sigma$.
\begin{figure}
    \centering
    \includegraphics[width=\textwidth]{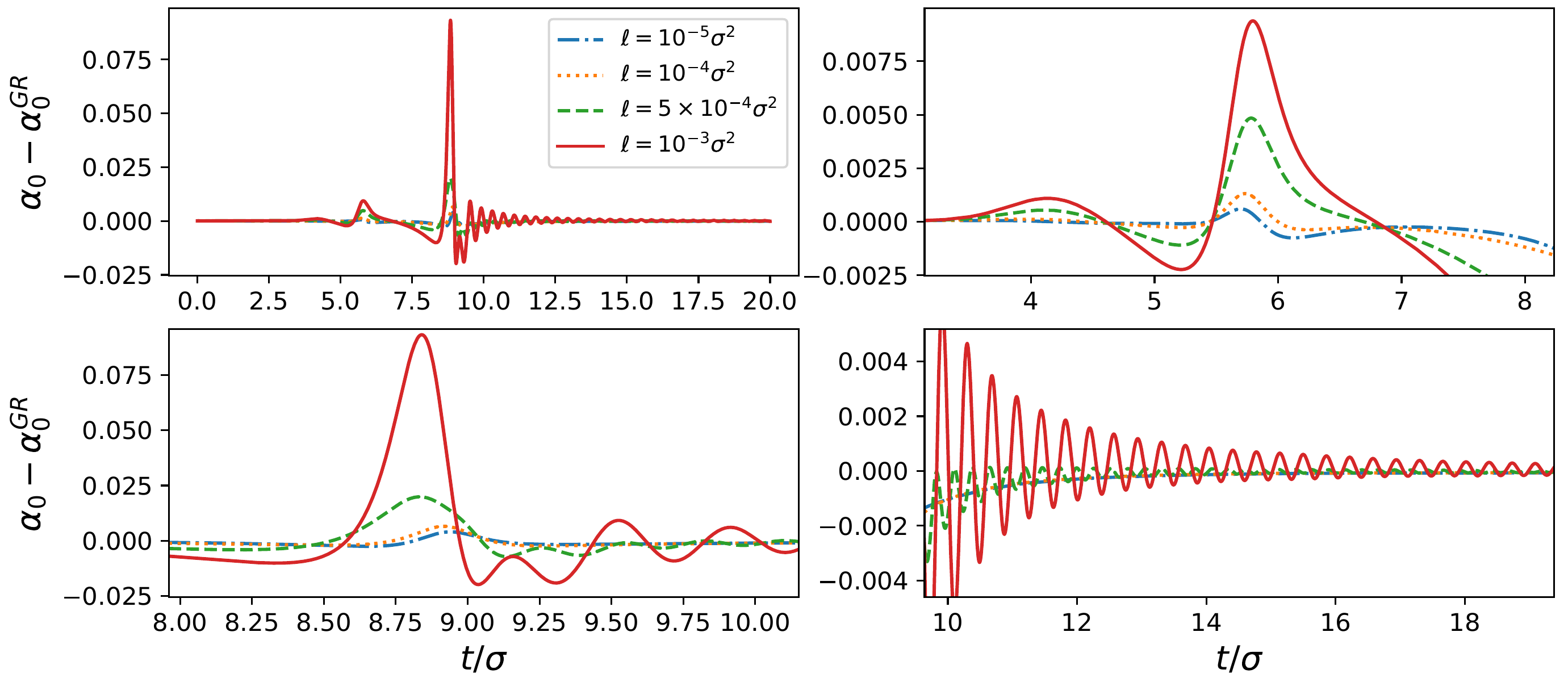}
    \caption{Plots of the central value for the lapse function for various values of $\ell$ after subtracting off that of GR ($\ell=0$). The dash-dotted blue, dotted orange, dashed green and solid red curves correspond to $\ell/\sigma^2=10^{-5}$, $10^{-4}$, $5\times10^{-4}$ and $10^{-3}$ respectively. The top-left panel shows the plots from $t=0$ until $t=20\sigma$ while the remaining panels show zoomed-in pieces of these plots. To produce these plots, we have set $\Delta r=0.01\sigma$, $\Delta t=\Delta r/2$ while taking the initial amplitude of the KG field to be $p=0.04$. The profiles converge to zero as $\ell$ decreases and we note that damped oscillations appear in the cases where $\ell/\sigma^2=5\times10^{-4}$ and $\ell/\sigma^2=10^{-3}$; with the amplitudes associated with the latter being greater in magnitude.}
    \label{fig:lapse_subtract_GR_multiple}
\end{figure}

Let us now study the deviation of the results for $\ell\neq0$ from the observed behaviour in the GR case. That is, we wish to analyse how the central value for the lapse function is affected by the non-GR contributions of a Starobinsky $f(R)$ model. In order to do this, we consider the cases where $\ell/\sigma^2=10^{-5}$, $10^{-4}$, $5\times10^{-4}$ and $10^{-3}$ and plot the central value  of the lapse function for each value of $\ell$, dubbed $\alpha_0$, after subtracting the central value for the lapse in the case of GR, dubbed $\alpha^{\text{GR}}_0$. These are shown in Figure \ref{fig:lapse_subtract_GR_multiple}. The top-left panel shows $\alpha_0-\alpha^{\text{GR}}_0$ from $t=0$ until $t=20\sigma$. The remaining three panels show zoomed-in pieces of the plots contained in the aforementioned top-left panel. The dash-dotted blue, dotted orange, dashed green and solid red curves correspond to $\ell/\sigma^2=10^{-5}$, $10^{-4}$, $5\times10^{-4}$ and $10^{-3}$ respectively. It is clear from Figure \ref{fig:lapse_subtract_GR_multiple} that, as the value for $\ell$ is decreased, the profiles approach zero as expected. For the cases where $\ell/\sigma^2=5\times10^{-4}$ and $10^{-3}$, damped oscillations starting around $t\sim10\sigma$ are seen while no such oscillations are clearly seen for the smaller values of $\ell/\sigma^2=10^{-5}$ and $\ell/\sigma^2=10^{-4}$. It is possible that similar oscillations may appear in the cases of $\ell/\sigma^2=10^{-5}$ and $\ell/\sigma^2=10^{-4}$ for larger subcritical values of $p$, however, this would require further investigation which is not covered in this work.

\section{Conclusions}\label{sec:conclusions}
We have for the first time generalized the BSSN formulation of the $f(R)$ gravity presented in \cite{Mongwane:2016qtz} to the case where the determinant of the conformal 3-metric is not necessarily unity, in particular for application to models in spherical symmetry. This modification, which was suggested in \cite{brown,brown2} for the case of GR, is referred to as the GBSSN formulation. The extension of this formalism to $f(R)$ theories of gravity in the metric formalism has allowed us to construct a numerical relativity code that we have applied to a system of one minimally coupled KG scalar field together with the Starobinsky $f(R)$ gravity. We note that, while the GBSSN formalism discussed here is for a general $f(R)$ theory, the aforesaid numerical relativity code is specific to the quadratic Starobinsky model. The latter leads to the usual Einstein term plus a quadratic contribution in the Ricci scalar which in recent years has attracted considerable attention in both cosmological and astrophysical studies.
This system is of particular interest since it can be easily generalized, on the one hand, to other scalar-tensor higher-order gravity theories, and on the other hand, to massive scalar fields with non-minimal couplings.

The formalism and its numerical implementation was verified through a number of standard numerical tests in the case of GR, demonstrating consistent and convergent results comparable to those found in the literature. In the case of a Schwarzschild black hole, previously used as a test case in~\cite{alcubierre,brown,Montero:2012yr,Baumgarte:2012xy}, we determined an apparent horizon mass which remained within $0.2\%$ of its expected value for the duration of the evolution. As a second test, we evolved a gauge pulse in order to induce dynamics in the metric components of a flat space-time~\cite{alcubierre,Montero:2012yr}. For this test, we used a partially implicit (PIRK) scheme following~\cite{Montero:2012yr} and demonstrated second-order convergence of the implementation.

Finally, we performed tests of the full SKG system, which in addition to the standard GBSSN variables also incorporates the Ricci scalar and a new variable, $W$, defined in equation~\eqref{eq:W_definition}. The PIRK treatment of the GBSSN equations was extended to accommodate these additional variables. As a diagnostic, we examined the central value for the lapse function as an indication of the strength of the local field, finding that a larger gravitational $R^2$-contribution lead to larger variations in this value. For a scalar field initially positioned at $r/\sigma=5$, damped oscillations begin at around $t\sim10\sigma$ in a subcritical scenario. The amplitude for these oscillations are greater when increasing the $R^2$-contribution in the gravitational action. The appearance of oscillatory behaviours in the evolution of the Starobinsky model, as well as in other power-law $f(R)$ models, has been reported in several contexts, ranging from the exterior solutions for realistic static neutron stars~\cite{AparicioResco:2016xcm} to the evolution of the matter power spectrum~\cite{Ananda:2008gs}. In many of them, the effect has been understood as a competing balance between the so-called {\it curvature effective fluid}, as generated by non-GR $f(R)$ terms, and the {\it standard matter fluid}, the latter usually divided by $f_R\neq0$ (see for instance this decomposition in \cite{Carloni:2004kp}, Sec. 2). When the field equations are adequately rewritten, the addition of those two fluids is covariantly conserved,  whereas separately none of them are.

Although the GBSSN formulation of $f(R)$ gravity can be applied to a variety of $f(R)$ models satisfying $f_{RR}\neq0$, we have only considered the Starobinsky gravity theory here given its importance and relevance in strong-gravity regimes. We have shown that the PIRK scheme with $R$ being treated explicitly while treating $W$ partially implicitly allows for the stable evolution of the SKG system. Nonetheless, this explicit-implicit approach may not be convenient to study the evolution of a KG scalar field in the presence of other $f(R)$ models. Therefore, a possible extension of this work would be to consider $f(R)$ models with a non-negligible contribution in the strong-gravity regime and see if the same PIRK scheme would allow for stable evolution. Complementarily, our code could be directly used to study the phenomenon of critical collapse for Starobinsky gravity using the GBSSN formulation. We also note that we have only considered the $f(R)$ gravity here when discussing the GBSSN formulation. Therefore, one would need to construct modifications to this formulation in order to consider more general higher-derivative gravitational theories. Nevertheless, the fact that we had to treat one of the additional fields compared to GR in the SKG system partially-implicitly may indicate that any numerical implementation of the GBSSN equations for other higher-derivative theories using the PIRK scheme may require some partially-implicit treatment of new variables.

\section*{Acknowledgements}
UKBV acknowledges financial support from the National Research Foundation (NRF) of South Africa (Grant number: PMDS22063029733), from the University of Cape Town Postgraduate Funding Office and from the Erasmus+ KA107 International Credit Mobility Programme. UKBV also acknowledges the hospitality of the IFT/UAM-CSIC Madrid (Spain) during the completion of the manuscript. AdlCD acknowledges financial support from South African NRF grants no.120390, reference: BSFP190416431035; no.120396, reference: CSRP190405427545; Grant PID2021-122938NB-I00 funded by MCIN/AEI/
10.13039/501100011033 and by ERDF - {\it A way of making Europe} and BG20/00236 action (Ministerio de Universidades, Spain). UKBV and AdlCD thank the hospitality of the Institute of Theoretical Astrophysics - University of Oslo (Norway) during the later stages of the preparation of the manuscript. The numerical simulations presented here were performed using the C++ programming language while all plots were produced using the Python programming language. All computations were performed on a consumer-grade laptop computer.

\appendix

\section{PIRK scheme for gauge pulse evolution}\label{sec:appendix_PIRK_gauge}
This appendix outlines the PIRK method used in this paper to obtain the stable evolution of a gauge pulse in flat space. While we use a vanishing shift vector for our evolutions, we state the scheme for an arbitrary value of $\mu$ in equation \eqref{eq:evolution_B} which includes the choices of a vanishing shift vector and the Gamma-driver condition.

Define $F_\alpha$, $F_\beta$, $F_a$, $F_b$, $F_\chi$, $F_K$, $F_A$, $F_{\bar\Delta^r}$ and $F_b$ as being the right-hand sides of equations \eqref{eq:evolution_lapse}, \eqref{eq:evolution_shift}, \eqref{eq:evolution_a}, \eqref{eq:evolution_b}, \eqref{eq:evolution_equation_chi}, \eqref{eq:evolution_trace_extrinsic_sph}, \eqref{eq:evolution_A_a}, \eqref{eq:evolution_Delta} and \eqref{eq:evolution_B}, respectively. In evolving the GBSSN variables, we evolve the 3-metric components explicitly while the other variables are evolved partially implicitly. For the latter, we are required to split the right-hand sides of the evolution equations into two parts: one, which will be treated explicitly and, the other one, to be treated partially implicitly. The evolution equations to be split are those associated with $K$, $A_a$ and $\bar\Delta^r$. The evolution equation for the auxiliary field $B^r$ has no part needing to be treated in an explicit way and therefore does not need to be split. However, if we were to include a damping factor, then it would be necessary to split the evolution equation.

The evolution equations for the extrinsic curvature terms, $K$ and $A_a$, as well as the regularized conformal connection function $\bar\Delta^r$ are split according to:
\begin{align}
F_{K,2}&:=\alpha\left(\frac32A_a^2+\frac{K^2}3\right)+\beta^rK'\ ,\\
F_{A,2}&:=\alpha KA_a+\beta^rA_a'+\frac{2\alpha\chi^2}{3a}\left(a\bar\Delta^r\phantom{}'+2a'\bar\Delta^r\right)\ ,\\
F_{\bar\Delta^r,2}&:=\beta^r\bar\Delta^r\phantom{}'\ ,
\end{align}
which are treated explicitly and
\begin{align}
F_{K,1}:=F_K-F_{K,2}\ ,\\
F_{A,1}:=F_A-F_{A,2}\ ,\\
F_{\bar\Delta^r,1}:=F_{\bar\Delta^r}-f_{\bar\Delta^r,2}\ ,
\end{align}
which are treated implicitly.

Suppose now that we are given the spatial profiles of the GBSSN variables at the $n^{\text{th}}$ time-step, $t_n$. That is, we have the data $\mathcal{D}_n:=\left\{\alpha_n,\beta^r_n,a_n,b_n,\chi_n,K_n,A_{a,n},\bar\Delta^r_n,B^r_n\right\}$ where for each $u_n\in\mathcal{D}_n$ we have written $u_n:=u(t_n,r)$. To describe the PIRK scheme, we split the procedure into eight steps which we discuss in some detail below.

\subsubsection*{Step 1.}
The first step in the PIRK procedure to evolve the GBSSN variables is to calculate the following quantities by using the data in $\mathcal{D}_n$ corresponding to the 3-metric components
\begin{align}
\tilde\alpha_n&=\alpha_n+\Delta tF_\alpha\left(\alpha_n,K_n\right)\ ,\\
\tilde\beta_n&=\beta^r_n+\Delta tF_\beta\left(B^r_n\right)\ ,\\
\tilde a_n&=a_n+\Delta tF_a\left(\beta^r_n,a_n,b_n,A_{a,n}\right)\ ,\\
\tilde b_n&=b_n+\Delta tF_b\left(\beta^r_n,a_n,b_n,A_{a,n}\right)\ ,\\
\tilde\chi_n&=\chi_n+\Delta tF_\chi\left(\alpha_n,\beta^r_n,a_n,b_n,\chi_n,K_n\right)\ .
\end{align}

\subsubsection*{Step 2.}
Next we use the extrinsic curvature data in $\mathcal{D}_n$ as well as the results from {\it Step 1} to compute
\begin{align}
\tilde K_n=K_n+\Delta t\left[\frac12F_{K,1}\left(\alpha_n,a_n,b_n,\chi_n\right)+\frac12F_{K,1}\left(\tilde\alpha_n,\tilde a_n,\tilde b_n,\tilde\chi_n\right)+F_{K,2}\left(\alpha_n,\beta^r_n,K_n,A_{a,n}\right)\right]\ ,
\end{align}
and
\begin{align}
\tilde A_{a,n}=A_{a,n}+\Delta t\bigg[\frac12F_{A,1}\left(\alpha_n,a_n,b_n,\chi_n\right)+\frac12F_{A,1}\left(\tilde\alpha_n,\tilde a_n,\tilde b_n,\tilde\chi_n\right)\nonumber\\
+F_{A,2}\left(\alpha_n,\beta^r_n,a_n,\chi_n,K_n,A_{a,n},\bar\Delta^r_n\right)\bigg]\ .
\end{align}

\subsubsection*{Step 3.}
We calculate
\begin{align}
\tilde{\bar\Delta}\phantom{}^r_{a,n}=\bar\Delta^r_{a,n}&+\Delta t\bigg[\frac12F_{\bar\Delta^r,1}\left(\alpha_n,\beta^r_n,a_n,b_n,\chi_n,K_n,A_{a,n}\right)\nonumber\\
&+\frac12F_{\bar\Delta^r,1}\left(\tilde\alpha_n,\tilde\beta^r_n,\tilde a_n,\tilde b_n,\tilde\chi_n,\tilde K_n,\tilde A_{a,n}\right)+F_{\bar\Delta^r,2}\left(\beta^r_n,\bar\Delta^r_n\right)\bigg]\ .
\end{align}

\subsubsection*{Step 4.}
We then calculate
\begin{align}
\tilde B^r_n=B^r_n+\frac{\Delta t}{2}\left[F_{B}\left(\bar\Delta^r_n\right)+F_{B}\left(\tilde{\bar\Delta}\phantom{}^r_n\right)\right]\ .
\end{align}

\subsubsection*{Step 5.}
After completing {\it Steps 1-4}, we then proceed to update the GBSSN variables. We start by updating the 3-metric components:
\begin{align}
\alpha_{n+1}&=\frac12\left[\alpha_n+\tilde\alpha_n+\Delta tF_\alpha\left(\tilde\alpha_n,\tilde K_n\right)\right]\ ,\\
\beta_{n+1}&=\frac12\left[\beta^r_n+\tilde\beta^r_n+\Delta tF_\beta\left(\tilde B^r_n\right)\right]\ ,\\
a_{n+1}&=\frac12\left[a_n+\tilde a_n+\Delta tF_a\left(\tilde \beta^r_n,\tilde a_n,\tilde b_n,\tilde A_{a,n}\right)\right]\ ,\\
b_{n+1}&=\frac12\left[b_n+\tilde b_n+\Delta tF_b\left(\tilde\beta^r_n,\tilde a_n,\tilde b_n,\tilde A_{a,n}\right)\right]\ ,\\
\chi_{n+1}&=\frac12\left[\chi_n+\tilde\chi_n+\Delta tF_\chi\left(\tilde\alpha_n,\tilde\beta^r_n,\tilde a_n,\tilde b_n,\tilde\chi_n,\tilde K_n\right)\right]\ .
\end{align}

\subsubsection*{Step 6.}
We then update the extrinsic curvature terms
\begin{align}
K_{n+1}=K_n+\frac{\Delta t}{2}\bigg[F_{K,1}\left(\alpha_n,a_n,b_n,\chi_n\right)+F_{K,1}\left(\alpha_{n+1},a_{n+1},b_{n+1},\chi_{n+1}\right)\nonumber\\
+F_{K,2}\left(\alpha_n,\beta^r_n,K_n,A_{a,n}\right)+F_{K,2}\left(\tilde\alpha_n,\tilde\beta^r_n,\tilde K_n,\tilde A_{a,n}\right)\bigg]\ ,
\end{align}
and
\begin{align}
A_{a,n+1}&=A_{a,n}+\frac{\Delta t}{2}\bigg[F_{A,1}\left(\alpha_n,a_n,b_n,\chi_n\right)+F_{A,1}\left(\alpha_{n+1},a_{n+1},b_{n+1},\chi_{n+1}\right)\nonumber\\
&+F_{A,2}\left(\alpha_n,\beta^r_n,a_n,\chi_n,K_n,A_{a,n},\bar\Delta^r_n\right)+F_{A,2}\left(\tilde\alpha_n,\tilde\beta^r_n,\tilde a_n,\tilde\chi_n,\tilde K_n,\tilde A_{a,n},\tilde{\bar\Delta}\phantom{}^r_n\right)\bigg]\ .
\end{align}

\subsubsection*{Step 7.}
Following the update of the extrinsic curvature, we update the regularized conformal connection function
\begin{align}
&\bar\Delta^r_{a,n+1}=\bar\Delta^r_{a,n}+\frac{\Delta t}{2}\bigg[F_{\bar\Delta^r,1}\left(\alpha_n,\beta^r_n,a_n,b_n,\chi_n,K_n,A_{a,n}\right)\nonumber\\
&+F_{\bar\Delta^r,1}\left(\alpha_{n+1},\beta^r_{n+1},a_{n+1},b_{n+1},\chi_{n+1},K_{n+1},A_{a,n+1}\right)+F_{\bar\Delta^r,2}\left(\beta^r_n,\bar\Delta^r_n\right)+F_{\bar\Delta^r,2}\left(\tilde\beta^r_n,\tilde{\bar\Delta}\phantom{}^r_n\right)\bigg]\ .
\end{align}

\subsubsection*{Step 8.}
Lastly, we update the auxiliary field
\begin{align}
B^r_{n+1}&=B^r_n+\frac{\Delta t}{2}\left[F_{B}\left(\bar\Delta^r_n\right)+F_{B}\left(\bar\Delta^r_{n+1}\right)\right]\ .
\end{align}

\section{PIRK scheme for GR with a KG scalar field}\label{sec:PIRK_matter_detailed}
In this section, we specify how we evolve the GBSSN equations numerically for the case of GR with a massless KG scalar field. More specifically, we indicate where in the PIRK scheme discussed in \ref{sec:appendix_PIRK_gauge} the three matter fields $\Phi$, $\Psi$ and $\Pi$ are introduced. Let us define $F_{\Phi}$, $F_{\Psi}$ and $F_{\Pi}$ as being the right-hand sides of the evolution equations \eqref{eq:evolve_Phi}, \eqref{eq:evolve_Psi} and \eqref{eq:evolve_Pi} respectively. We treat $\Phi$ and $\Psi$ explicitly, while $\Pi$ is treated partially implicitly. In order to treat $\Pi$ partially implicitly, we need to split the evolution equation into two parts: the first, which will be treated explicitly and, a second, which will carry the implicit treatment. To this end, we define
\begin{align}
F_{\Pi,2}:=\alpha K\Pi+\beta^r\Pi'\ ,
\end{align}
and further define
\begin{align}
F_{\Pi,1}:=F_{\Pi}-F_{\Pi,2}\ .
\end{align}
With the above definitions in mind, let us turn our attention to implementing the evolution of the matter fields in the PIRK scheme given in \ref{sec:appendix_PIRK_gauge}. Below, we refer to the steps stated therein.

To introduce a massless KG scalar field into the scheme presented in \ref{sec:appendix_PIRK_gauge}, we first compute the following in {\it Step 1} of the scheme
\begin{align}
\tilde\Phi_n&=\Phi_n+\Delta tF_{\Phi}\left(\alpha_n,\beta^r_n,\Psi_n,\Pi_n\right)\,,\\
\tilde\Psi_n&=\Psi_n+\Delta tF_{\Phi}\left(\alpha_n,\beta^r_n,\Psi_n,\Pi_n\right)\,.
\end{align}
For the field $\Pi$, we compute the following after {\it Step 2} and before {\it Step 3}
\begin{align}
\tilde\Pi_n=\Pi_n+\Delta t\bigg[\frac12F_{\Pi,1}\left(\alpha_n,a_n,b_n,\chi_n,\Phi_n,\Psi_n\right)+\frac12F_{\Pi,1}\left(\tilde\alpha_n,\tilde a_n,\tilde b_n,\tilde\chi_n,\tilde\Phi_n,\tilde\Psi_n\right)\nonumber\\
+F_{\Pi,2}\left(\alpha_n,\beta^r_n,K_n,\Pi_n\right)\bigg]\,.
\end{align}
Following this, we then update the variables $\Phi$ and $\Psi$ in {\it Step 5}
\begin{align}
\Phi_{n+1}=\frac12\left[\Phi_n+\tilde\Phi_n+\Delta tF_{\Phi}\left(\alpha_n,\beta^r_n,\Psi_n,\Pi_n\right)\right]\ ,
\end{align}
and
\begin{align}
\Psi_{n+1}=\frac12\left[\Psi_n+\tilde\Psi_n+\Delta tF_{\Pi}\left(\alpha_n,\beta^r_n,\Psi_n,\Pi_n\right)\right]\ .
\end{align}
Lastly, we then update $\Pi$ using
\begin{align}
\Pi_{n+1}=\Pi_n+\frac{\Delta t}{2}\bigg[&F_{\Pi,1}\left(\alpha_n,a_n,b_n,\chi_n,\Phi_n,\Psi_n\right)+F_{\Pi,1}\left(\alpha_{n+1},a_{n+1},b_{n+1},\chi_{n+1},\Phi_{n+1},\Psi_{n+1}\right)\nonumber\\
&+F_{\Pi,2}\left(\alpha_n,\beta^r_n,K_n,\Pi_n\right)+F_{\Pi,2}\left(\tilde\alpha_n,\tilde\beta^r_n,\tilde K_n,\tilde\Pi_n\right)\bigg]\ ,
\end{align}
which is implemented after {\it Step 6} and before {\it Step 7}. It is worth noting that, when updating the extrinsic curvature, all terms containing $\Phi$ and $\Psi$ are contained in $F_{A,1}$ and $F_{K,1}$ while all terms depending on $\Pi$ are contained in $F_{K,2}$ and do not appear in $F_{A,2}$. On the other hand, for the treatment of the regularized conformal connection function, all terms depending on the relevant matter fields are contained in $F_{\bar\Delta\phantom{}^r,1}$.
\section{PIRK with $f(R)$ contributions}\label{prik_f_R_appendix}
In this section, we specify how the $f(R)$ $(f_{RR}\neq0)$ GBSSN variables $R$ and $W$ are introduced into the PIRK scheme presented in \ref{sec:appendix_PIRK_gauge}. As mentioned in Subsection \ref{sec:numerical_f_R}, we treat $R$ explicitly while treating $W$ partially implicitly. Let us define $F_W$ and $F_R$ as being the right-hand sides of equations \eqref{eq:evolve_W_GBSSN} and \eqref{eq:evolve_R_GBSSN} respectively. In addition, we define
\begin{align}
F_{W,2}&:=\alpha KW+\beta^rW'\,,\\
F_{W,1}&:=F_W-F_{W,2}\,.
\end{align}

Since $R$ is treated explicitly, we evolve it together with the metric components and, when including a massless KG scalar field, with $\Phi$ and $\Psi$. That is, in {\it Step 1} of the scheme given in \ref{sec:appendix_PIRK_gauge} we compute
\begin{align}
\tilde R_n=R_n+\Delta tF_R\left(\alpha_n,\beta^r_n,R_n,W_n\right)\ .
\end{align}
For $W$, we compute the following in {\it Step 2}
\begin{align}
\tilde W_n=W_n+\Delta t\bigg[\frac12F_{W,1}\left(\alpha_n,a_n,b_n,\chi_n,R_n\right)+\frac12F_{W,1}\left(\tilde\alpha_n,\tilde a_n,\tilde b_n,\tilde\chi_n,\tilde R_n\right)\nonumber\\
+F_{W,2}\left(\alpha_n,\beta^r_n,K_n,W_n\right)\bigg]\ .
\end{align}
We then update the Ricci scalar in {\it Step 5} as follows
\begin{align}
R_{n+1}=\frac12\left[R_n+\tilde R_n+\Delta tF_R\left(\tilde\alpha_n,\tilde\beta^r_n,\tilde R_n,\tilde W_n\right)\right]\ .
\end{align}
Lastly, we update $W$ in {\it Step 6} by computing
\begin{align}
W_{n+1}=W_n+\frac{\Delta t}{2}\bigg[F_{W,1}\left(\alpha_n,a_n,b_n,\chi_n,R_n\right)+F_{W,1}\left(\alpha_{n+1},a_{n+1},b_{n+1},\chi_{n+1},R_{n+1}\right)\nonumber\\
+F_{W,2}\left(\alpha_n,\beta^r_n,K_n,W_n\right)+F_{W,2}\left(\tilde\alpha_n,\tilde\beta^r_n,\tilde K_n,\tilde W_n\right)\bigg]\ .
\end{align}
It now remains to state how we treat the terms containing $R$ and $W$ when updating the extrinsic curvature and the regularized conformal connection function. For $K$ and $A_a$, we place the terms containing $R$ into $F_{K,1}$ and $F_{A,1}$ while there is no dependence on $W$. For $\bar\Delta\phantom{}^r$, all terms depending on $R$ and $W$ are placed in $F_{\bar\Delta\phantom{}^r,1}$.
\newpage
\bibliography{references}
\end{document}